\documentclass[sn-standardnature]{sn-jnl}% Standard Nature Portfolio Reference Style
\jyear{2021}%

\theoremstyle{thmstyleone}%
\theoremstyle{thmstyletwo}%
\usepackage{amsmath}
\theoremstyle{thmstylethree}%
\usepackage{xcolor}
\usepackage{natbib}

\begin{document}

%\title{The `Evolving' Story of a baby Sub-Neptune/Super-Earth: V1298 Tau b}

\title{The metal-poor atmosphere of a Neptune/Sub-Neptune planet's progenitor}

%\title{The low-metallicity spectrum of a young Sub-Neptune/Super-Earth progenitor}

%\author[0000-0002-3522-5846]
\author*[1]{\fnm{Saugata} \sur{Barat}}\email{s.barat@uva.nl}

\author[1]{\fnm{Jean-Michel} \sur{D\'esert}}\email{desert@uva.nl}

\author[2]{\fnm{Allona} \sur{Vazan}}\email{vazan@openu.ac.il}

\author[1]{\fnm{Robin} \sur{Baeyens}}\email{r.l.l.baeyens@uva.nl}


\author[3]{\fnm{Michael R.} \sur{Line}}\email{mrline@asu.edu}

\author[4]{\fnm{Jonathan J.} \sur{Fortney}}\email{jfortney@ucsc.edu}

\author[5]{\fnm{Trevor J.} \sur{David}}\email{tdavid@flatironinstitute.org}

\author[6,7,8]{\fnm{John H.} \sur{Livingston}}\email{john.livingston@nao.ac.jp}
\author[1]{\fnm{Bob} \sur{Jacobs}}\email{b.p.j.jacobs@uva.nl}
\author[1,9,12]{\fnm{Vatsal} \sur{Panwar}}\email{vatsal.panwar@warwick.ac.uk}
%\author[1]{\fnm{James} \sur{Sikora}}\email{j.t.sikora@uva.nl}
\author[1]{\fnm{Hinna} \sur{Shivkumar}}\email{h.shivkumar@uva.nl}
\author[1]{\fnm{Kamen O.} \sur{Todorov}}\email{kamen.o.todorov@gmail.com}
\author[10]{\fnm{Lorenzo} \sur{Pino}}\email{lorenzo.pino@inaf.it}
\author[1]{\fnm{Georgia} \sur{Mraz}}\email{georgiamraz@gmail.com}
\author[11]{\fnm{Erik A.} \sur{Petigura}}\email{petigura@astro.ucla.edu}

\affil[1]{\orgdiv{Anton Pannekoek Institute}, \orgname{University of Amsterdam}, \city{Amsterdam}, \postcode{1098XH},  \country{Netherlands}}
\affil[2]{\orgdiv{Astrophysics Reseach Center (ARCO)}, \orgname{The Open University of Israel}, \city{Ra’anana}, \postcode{43107},  \country{Israel}}
\affil[3]{\orgdiv{School of Earth and Space Exploration}, \orgname{Arizona State University}, \city{Tempe}, \postcode{ AZ 85287},  \country{USA}}
\affil[4]{\orgdiv{Department of Astronomy \& Astrophysics}, \orgname{ University of California}, \city{Santa Cruz}, \postcode{CA 95064},  \country{USA}}
\affil[5]{\orgdiv{Center for Computational Astrophysics}, \orgname{ Flatiron Institute}, \city{New York}, \postcode{ NY 10010},  \country{USA}}
\affil[6]{\orgdiv{Astrobiology Center}, \city{2-21-1 Osawa, Mitaka, Tokyo} \postcode{181-8588}, \country{Japan}}
\affil[7]{\orgdiv{National Astronomical Observatory of Japan},\city{2-21-1 Osawa, Mitaka, Tokyo} \postcode{181-8588}, \country{Japan}}
\affil[8]{\orgdiv{Department of Astronomy}, \orgname{Graduate University for Advanced
Studies (SOKENDAI)}, \city{Tokyo}, \postcode{ 2-21-1 Osawa, Mitaka},  \country{Japan}}
\affil[9]{\orgdiv{Department of Physics}, \orgname{University of Warwick}, \city{Coventry} \postcode{CV4 7AL}, \country{UK}}
\affil[10]{\orgdiv{INAF - Osservatorio Astrofisico di Arcetri}, \orgname{ Largo E. Fermi 5}, \city{Firenze}, \postcode{   50125},  \country{Italy}}
\affil[11]{\orgdiv{Department of Physics \& Astronomy}, \orgname{  University of California Los Angeles}, \city{Los Angeles}, \postcode{ CA, 90095},  \country{USA}}
\affil[12]{\orgdiv{Centre for Exoplanets and Habitability}, \orgname{University of Warwick}, \city{Coventry} \postcode{CV4 7AL}, \country{UK}}


\abstract{

Young transiting exoplanets offer a unique opportunity to characterize the atmospheres of fresh and evolving products of planet formation. We present the transmission spectrum of V1298 Tau b; a 23 Myr old warm Jovian sized planet orbiting a pre-main sequence star. We detect a primordial atmosphere with an exceptionally large atmospheric scale height and a water vapour absorption at 5$\sigma$
level of significance.  We estimate a mass and density upper limit (24$\pm$5$M_{\oplus}$, 0.12gm/$cm^{3}$ respectively). V1298 Tau b is one of the lowest density planets discovered till date. We retrieve a low atmospheric metallicity (logZ=$-0.1^{+0.66}_{-0.72}$ solar), consistent with solar/sub-solar values. Our findings challenge the expected mass-metallicity from core-accretion theory. Our observations can be explained by in-situ formation via pebble accretion together with ongoing evolutionary mechanisms.
We do not detect methane, which hints towards a hotter than expected interior from just the formation entropy of this planet. Our observations suggest that V1298 Tau b is likely to evolve into a Neptune/sub-Neptune type of planet.

%Young transiting exoplanets offer a unique opportunity to characterize the atmospheres of fresh and evolving products of planet formation. We present the transmission spectrum of V1298 Tau b; a 23 Myr old low density warm Jovian sized planet orbiting a pre-main sequence star. We detect a mostly clear primordial atmosphere with an exceptionally large atmospheric scale height and a water vapour absorption detected at 5 $\sigma$ level of significance. We estimate an upper limit for the planet mass (24$\pm$ 5$M_{\oplus}$) from the observed spectrum. We retrieve solar/sub-solar atmospheric metallicity for V1298 Tau b, challenging planet formation theory predictions for Neptune mass planets. We do not detect methane, which hint towards a hotter interior than expected from just the formation entropy of this young Neptune-mass planet. Our observations suggest that V1298 Tau b will likely evolve to become a sub-Neptune or a super-Earth type of planet.
}


\maketitle

Exoplanet population studies reveal the crucial impact of planet formation and early evolutionary mechanisms \cite[E.g, see][]{petigura2013,Fulton_2017} on their demographic characteristics. However, evolutionary processes such as, atmospheric mass loss driven by host star XUV flux \cite[][]{owen2013}, interior cooling \cite[E.g, see][]{gupta2019}, and contraction \cite[E.g, see][]{kubyshkina2020} can significantly alter their thermal structure and composition within the first 100Myrs, thereby obscuring the imprints of planet formation. In this context, young transiting exoplanets represent a unique opportunity to probe the atmosphere of freshly formed planets and test formation and early evolution theories \cite{lee_chiang2016,lee2014,lee2019,owen_2020}. However, studying these young planets is challenging as most of them do not have well constrained masses, due to large uncertainties in radial velocity (RV) measurements from their highly variable host stars \cite{blunt2023}. Young stars are known to have large spot coverage and frequent flaring activity \cite{feinstein2021,feinstein2020}, which can contaminate the measured transmission spectrum due to the transit light source effect \cite{rackham_2019}.
Most of the known young transiting planets (E.g, see\cite{david2016,mann2022}) lie above the radius-valley \cite[e.g, see][]{benatti2021} and are theoretically predicted to be Neptune or sub-Neptune/super-Earth progenitors \cite{kubyshkina2020,Modirrousta-Galian2020}. 

The V1298 Tau is one of the youngest transiting multi-planet system known so far, consisting of 3 confirmed planets in a near 3:2:1 mean motion resonance and a fourth planet with an unconfirmed period \citep[][]{david_19_b,david19,sikora2023,Feinstein2022}.
The host is a 23Myr old weak-lined T-Tauri star, which is a member of Group 29; a young association in the foreground of the Taurus-Auriga star forming region \citep{oh17,luhman18}. Several age estimates have been published for V1298 Tau: 23$\pm$4Myr \citep{david19}, 20$\pm$10Myr \citep{mascareno_2021}, 28$\pm$4Myr \citep{johnson2022}. All these estimates agree within 1$\sigma$ and we adopt 23$\pm$4 from \citep{david_19_b}. We observed one primary transit of V1298 Tau b using 10 HST orbits with the WFC3/G141 instrument for Program GO 16083 (See Methods Observations). V1298 Tau b is a warm ($T_{eq}=670 K$) \cite{david_19_b}, Jovian sized planet (0.8-0.9$R_{J}$) \cite{sikora2023,Feinstein2022,david_19_b} orbiting its host star in 24.14 days \cite{david_19_b,david19,Feinstein2022,sikora2023}. Mass measurement using RVs report Jovian mass (0.64$\pm$0.19$M_{J}$ \cite{mascareno_2021}, $<$0.5$M_{J}$ \cite{sikora2023}), however the reliability of these constraints have been questioned recently \cite{blunt2023}. Using these mass measurements, \citep{maggio2022} concluded that V1298 Tau b would be stable to atmospheric mass loss due to its strong gravity.

%Details of the observations are outlined in Methods Observations. 

The raw HST images were reduced using a custom pipeline \citep[][]{jacobs2022} (See Methods Data Reduction for details). We extract a broadband integrated `white' light curve in the  HST/WFC3 G141 bandpass (1.12$\mu$m-1.65$\mu$m) and use a divide-white common mode approach to derive systematics-corrected spectroscopic light curves \cite{kreidberg14}. The extracted white and de-trended spectroscopic light curves are shown in Extended Data Figure \ref{fig:white_lc} and  \ref{fig:spectroscopic_lc} respectively. The de-trended spectroscopic light curves are fitted with a \texttt{batman} planetary transit model, linear limb darkening and a linear stellar baseline ( For details, see Methods Light curve analysis and Table \ref{tab:table2}). We estimate the effect of unocculted star spots on the transmission spectrum using techniques outlined in \cite[][]{rackham_2019} (See Methods Stellar activity).

\section*{Results} \label{section:results}

\begin{figure}
\centering
%\begin{tabular}

  \includegraphics[width=\linewidth]{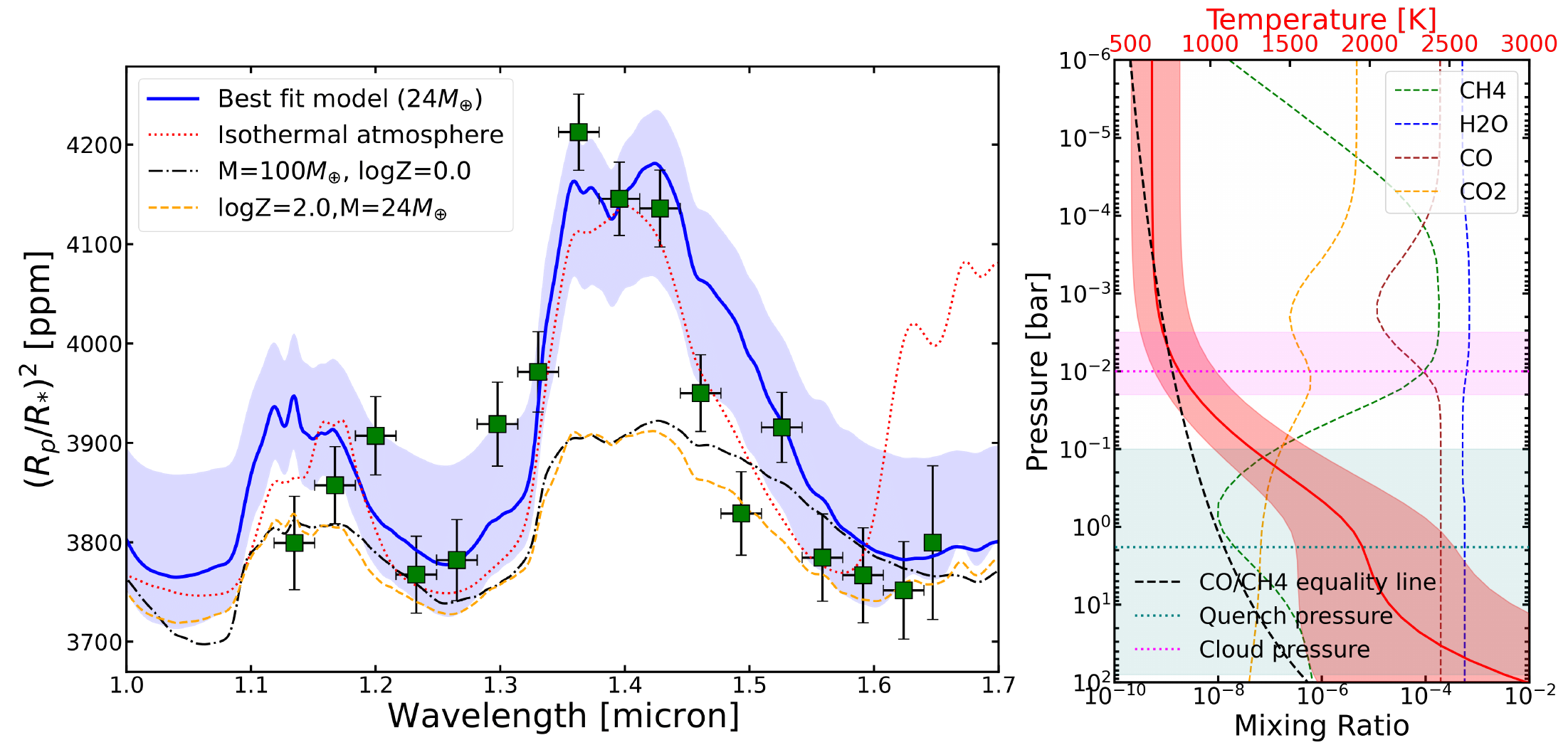}
 % \caption{Transmission spectrum of V1298 Tau b}
 % \label{fig:sfig1}
%\end{tabular}%
%\begin{tabular}

%  \includegraphics[width=0.35\linewidth]{images/V1298b_T-P+abundance_v3.pdf}
 % \caption{Retrieved T-P profile}
 % \label{fig:sfig2}
%\end{tabular}
 \caption{Left panel: Observed HST/WFC3 transmission spectrum (without stellar activity correction) of V1298 Tau b (green squares) with one-sigma error bars from which an upper limit of the planet mass is determined (24$M_{\oplus}$) . Atmospheric retrievals with the estimated mass upper limit show that the observations are consistent with solar/sub solar atmospheric metallicity (Solid blue line). The dash-dotted black line shows a transmission spectrum for a 100$M_{\oplus}$, solar metallicity model, and the orange line represents a 24$M_{\oplus}$ with a 100 times solar metallicity model. Both these models fail to capture the amplitude of the water feature and can be ruled out at $\sim 5\sigma$ confidence. The red dotted model represents an isothermal model, which shows an absorption feature around 1.6$\mu$m due to methane. An isothermal equilibrium chemistry model without high internal temperature and vertical mixing fails to explain the observed spectrum around 1.6$\mu$m (See Results section). Retrievals with lower masses have been explored in Methods Mass estimate (See Extended Data Figure \ref{fig:mass_uncertainty}). Stellar activity corrected transmission spectrum (See Methods Stellar Activity) for V1298 Tau b is consistent within 1$\sigma$ with observed uncorrected spectrum. Right panel: Retrieved T-P profile (24$M_{\oplus}$ model) with the 1$\sigma$ confidence interval (red shaded region). The dashed lines of different colours represent the equilibrium abundances for the chemical species included in our model (calculated for the red solid T-P profile). The magenta and blue dotted lines and the corresponding shaded region show the location of the retrieved grey cloud deck and quenching pressure from our retrieval analysis. (See Results section and Table \ref{tab:table3}). }

%\end{center}
\label{fig:fig1}
\end{figure}

\begin{figure}
\centering
%\begin{tabular}

  \includegraphics[width=0.75\textwidth]{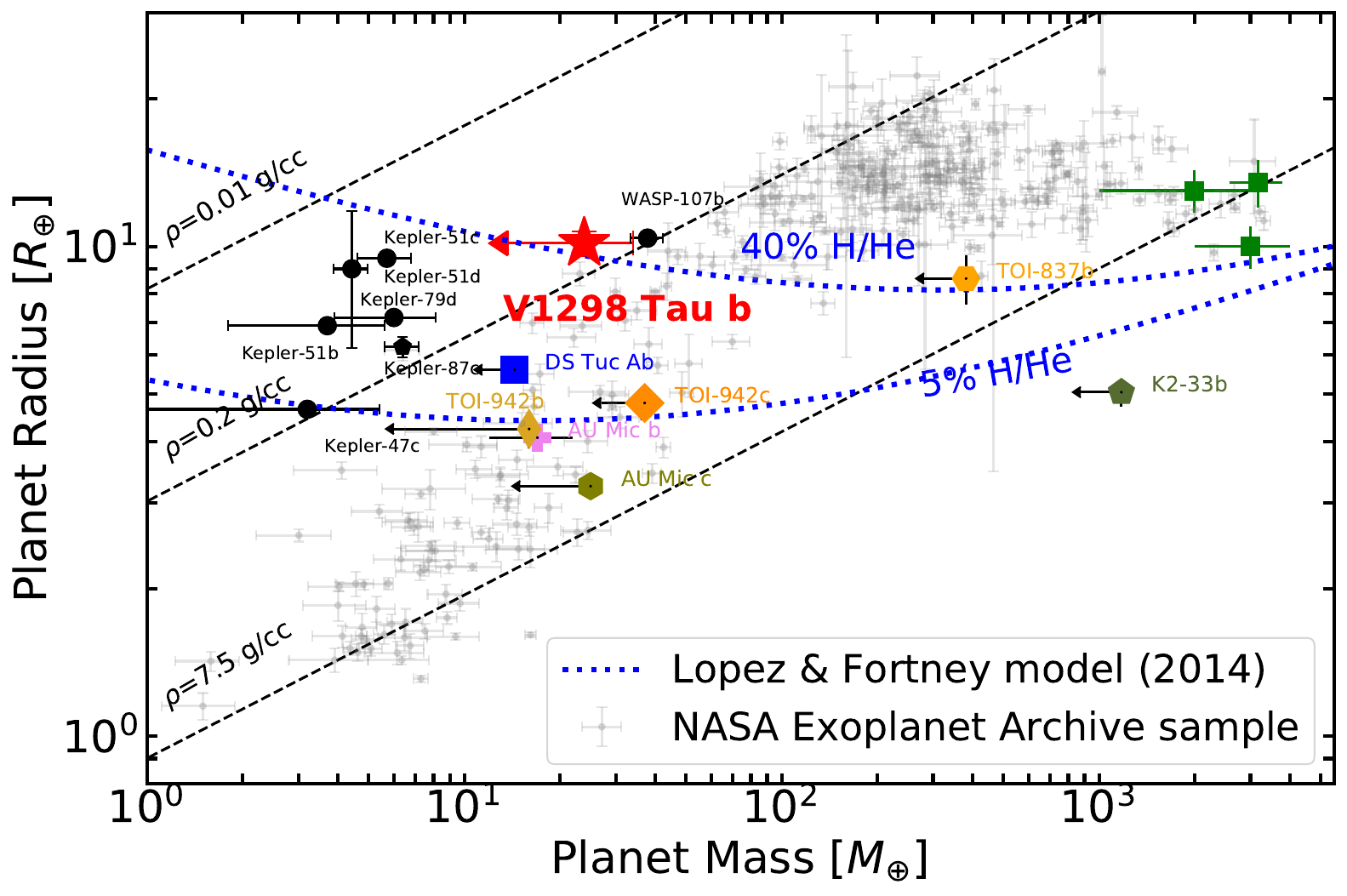}
 % \caption{Transmission spectrum of V1298 Tau b}
 % \label{fig:sfig1}
%  \end{tabular}%
  \vspace{0.1cm}
%\begin{tabular}

  \includegraphics[width=0.75\textwidth]{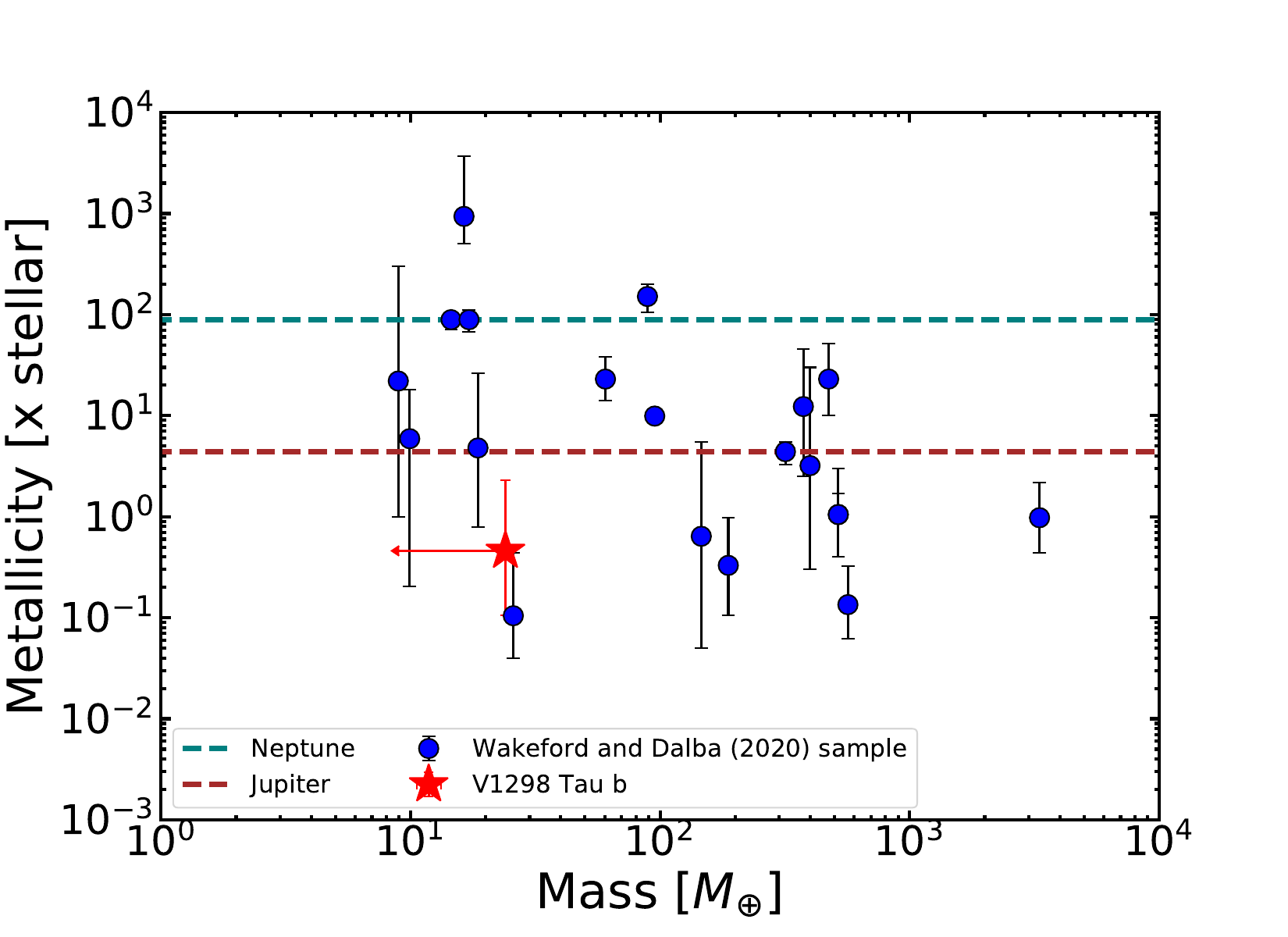}
 % \caption{Retrieved T-P profile}
 % \label{fig:sfig2}
%\end{tabular}
 \caption{ Upper panel: V1298 Tau b (Red star) in the mass-radius diagram; the mass upper limit was calculated from the observed transmission spectrum using the formalism presented in \cite{de_wit}. V1298 Tau b is shown in comparison with other known young planets and low-density `super-puffs'. Grey dots are known mature planets(obtained from NASA Exoplanet Archive). The blue dotted lines are theoretical models from \cite[][]{Lopez_2014}, and show that the measured mass and radius of V1298 Tau b is consistent with an atmosphere with a significant H/He envelope ($\sim$ 40$\%$ mass fraction assuming 10$M_{\oplus}$ core).  V1298 Tau b is amongst the lowest density (0.12gm/$cm^{3}$ planets discovered. Lower panel: V1298 Tau b (red star) shown in a Mass-metallicity diagram with a sample of exoplanets compiled from \cite{wakeford_dalba}. The atmospheric metallicity is derived from the retrieval analysis (See Results section). The dashed blue and brown lines show the metallicities of Neptune and Jupiter respectively. We note that the solar system metallicity estimates are from methane abundance measurements \cite{atreya2016}, whereas for exoplanets the metallicity estimates are derived from oxygen abundance measurement. V1298 Tau b has a mass consistent with Neptune/sub-Neptunes or potentially even super-Earths, but its metallicity is comparable to giant planets like Jupiter. }
 \label{fig:fig2}
\end{figure}

The transmission spectrum of V1298 Tau b (see Figure \ref{fig:fig1}) shows a high amplitude absorption feature around the 1.4$\mu$m water band ($\sim$ 400 ppm), which is the largest among the known Neptune/super-Neptune mass planets, such as HAT-P-26b \cite{wakeford_17} (250 ppm) and GJ 3470b (150 ppm) \cite{benneke_19}. The water absorption amplitude is large compared to well studied hot Jupiters, such as HD209458b ($\sim$200ppm, \cite{deming13})
The amplitude of the water feature is indicative of a large atmospheric scale height, revealing an extended H-rich atmosphere. Assuming a cloud free, H/He rich and isothermal atmosphere we constrain the scale height of this planet ($\sim$ 1000km), from which we estimate the mass to be 24 $\pm$ 5$M_{\oplus}$ assuming a clear atmosphere (see Methods, Mass estimate) using a known method \cite{de_wit}. This mass estimate becomes an upper limit if the atmosphere is partly cloudy or hazy. Radial velocity measurements of this system report Jovian/sub-Jovian mass (220 $\pm$ 70$M_{\oplus}$) for V1298 Tau b\cite[][]{mascareno_2021,sikora2023}. However, our observation rules out a 100$M_{\oplus}$ ($\sim 2\sigma$ lower limit from \cite{mascareno_2021}) transmission spectrum model (Figure \ref{fig:fig1}) at $\sim$ 5$\sigma$ confidence. 
We compare the derived mass and radius of planet b to the population of exoplanets (Figure \ref{fig:fig2}). V1298 Tau b, with a density upper limit of 0.12gm/$cm^{3}$, is comparable to lowest density planets known  (`super-puffs')\cite[E.g, see][]{jontoff-hunter2014}, however V1298 Tau b has a clear atmosphere compared to most super-puffs \cite[E.g see][]{libby_robert2020}. The estimated mass upper limit of V1298 Tau b is consistent with a Neptune/sub-Neptune mass planet with a substantial H/He envelope ($\sim$40$\%$, Figure \ref{fig:fig2}).

The observed transmission spectrum (without stellar activity correction) is modelled using a 1D radiative transfer code (\texttt{PetitRADTRANS}) \cite[][]{molliere_19}. We fix the planet mass to the estimated upper limit (24$M_{\oplus}$). We model the atmosphere using an analytic temperature-pressure profile \cite{guillot2010} and a gray cloud deck.  The dominant carbon bearing species at 670K is expected to be methane \cite{fortney_2020}. However, we do not detect methane absorption around 1.6$\mu$m (Figure \ref{fig:fig1}). Absence of methane has been reported for other warm planets \cite[E.g, see][]{benneke_19}. which can be potentially explained by vertical mixing \cite{fortney_2020}. We simulate the effect of vertical mixing using a `quench' pressure in our models (See Methods Atmospheric modelling). The observed and modelled transmission spectra, and the retrieved atmospheric properties are shown in Figure \ref{fig:fig1}. The retrievals converge to a low atmospheric metallicity (solar/sub-solar) compared to theoretical expectations from core-accretion \cite{pollack1996} and known constraints for sub-Neptunes/super-Earths \cite[E.g, see][]{moses2013,gao2023}. The observed spectrum can be equally explained by lower planet masses but even lower metallicities (0.1-0.01 solar). Cloudy models are statistically favoured to cloud free models (See Extended Data Figure \ref{fig:mass_uncertainty} and Table \ref{tab:table3}). From the derived mass upper limit (this work) and radius measurement from \citep{david_19_b}, we derive an upper limit of 6$\times$solar on the atmospheric metallicity of V1298 Tau b applying the formalism from \citep{thorngren2019}. This upper limit should be interpreted cautiously as the models presented in \citep{thorngren2019} do not account for high interior flux from the planet. Since V1298 Tau b is young with a potentially hot interior, the true bulk metallicity could be higher. However, the posteriors from our retrievals (Extended Data Figure \ref{fig:retrieval_posterior}) rule out 6$\times$solar values at $\sim3\sigma$, implying relatively unmixed interior -atmosphere structure for this planet.

We present V1298 Tau b in the context of the exoplanet population (Figure \ref{fig:fig2}; lower panel). V1298 Tau b has a mass consistent with a Neptune/sub-Neptune or even potentially a super-Earth and a metallicity comparable or lower to Jupiter. High metallicity atmosphere (100$\times$solar) with the estimated mass upper limit of this planet can be ruled out at $\sim 5\sigma$ confidence (See orange dashed model in Figure \ref{fig:fig1}). Therefore, in spite of being a likely Neptune/sub-Neptune, or even a super-Earth progenitor, V1298 Tau b possesses an atmosphere that is 100-1000 times metal depleted compared to Neptune and Uranus.

In addition, we studied the origin of the absence of expected methane. We performed chemical kinetics models which incorporate a self-consistent T-P profile and vertical mixing (see Methods Atmospheric models). These models with different internal temperatures demonstrate that it is possible to remove methane through deep quenching, although it requires a high interior temperature (see Fig.~\ref{fig:chem_tint_varied}). At the highest intrinsic temperature we have tested ($T_\textrm{int} = 400$~K), the quenched molar fraction of methane is still $10^{-4.7}$, which is close to the observability limit ($10^{-5.5}$) of HST for methane \citep{fortney_2020}. Retrievals using free chemistry (See Methods Atmospheric Modelling and Supplementary Information Figure 1) put an upper limit of $10^{-6}$ on the methane Volume Mixing Ratio (VMR).

\begin{figure}
    \centering
    \resizebox{9cm}{!}{\includegraphics[width=\linewidth]{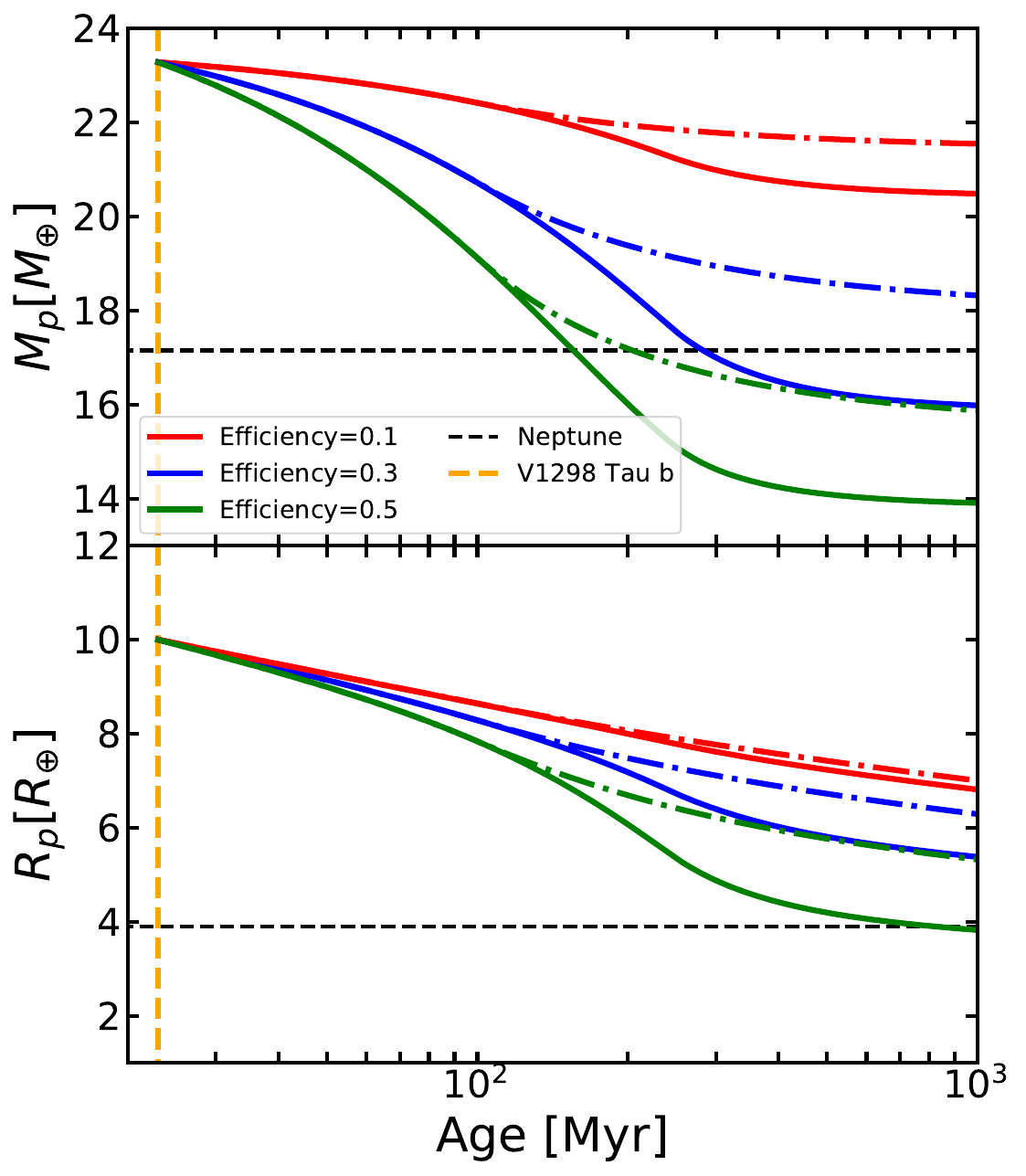}}
    \caption{Mass (upper panel) and radius (lower panel) evolutionary tracks simulated for V1298 Tau b during the first Gigayear using energy limited atmospheric evolution models presented in the \texttt{platypos} code \cite{poppenhaeger_21} (See Methods, Atmospheric evolution models).
    The radius evolution is a combined effect of atmospheric contraction and mass loss.
   Simulations for different values of the mass loss efficiency parameter (0.1,0.3,0.5) are shown with different colours. The solid lines show a high stellar activity track (Activity timescale 250Myr) and the dash-dot lines show a low stellar activity track (Activity timescale  100Myr, for details see \cite{poppenhaeger_21}). These models show that for moderate photoevaporation efficiency and high stellar activity, this planet is likely to lose mass and end up as a Neptune/sub-Neptune or even potentially a super-Earth depending on its mass.}
    
%    The code calculates the mass loss rate at a given point in time, using the energy-limited mass loss formalism \cite[e.g see][]{owen_jackson2012, salz2015} and evolve the planet's physical properties (mass and radius) at every step of the calculation. The radius evolution is a combined effect of atmospheric contraction and mass loss, and the updated size of the planets are calculated from the scaling relation given in \cite{Lopez_2014}. We perform the simulation for different values of the mass loss efficiency parameter (0.1,0.3,0.5) shown with different colours. The solid lines show a high stellar activity track (Activity timescale 250Myr) and the dash-dot lines show a low stellar activity track (Activity timescale  100Myr, for details see \cite{poppenhaeger_21}). We adopt the stellar luminosity from \cite{poppenhaeger_21}. These simulations have performed considering the estimated mass upper limit (24$M_{\oplus}$). For lower masses we can expect higher mass loss rates. 

    \label{fig:fig3}
\end{figure}

\section*{Discussion} \label{section:discussion}

The differences between the mass estimate from the atmospheric scale height and those from dynamical studies \citep{mascareno_2021,sikora2023} could potentially originate from the treatment of the impact of stellar activity on RV signals. The robustness of mass estimates from RVs of this system have been questioned recently \cite{blunt2023}. There is also uncertainty on the orbital period of planet e \citep{Feinstein2022} in this system which could significantly impact the RV mass constraints. Efforts to characterize the mass from transit timing variation measurements are ongoing (Livingston et. al, in prep).

The low envelope metallicity and relatively large H/He content that we measure for V1298 Tau b is in agreement with early evolution models \cite{Lopez_2014}, yet this planet must have been on the verge of runaway gas accretion. We emphasise that the origin and early evolution of Neptunes/sub-Neptunes has been an open question: it is unknown why these planets accreted only a small fraction of H/He and did not become gas giants \cite{lee2014,lee2019}. Such systems likely formed in-situ; either early with an enhanced atmospheric opacity due to dust grains \cite{lee2014}, or with significant disk-envelope interaction to replenish the proto-atmosphere with high entropy gas \cite{ormel2015}. Late formation in a depleting transitional disk such that the core does not have enough time to accrete a large amount of H/He envelope \cite{lee_chiang2016} can also produce Neptune/sub-Neptune mass planets. 

The standard core-accretion picture of planet formation \cite{pollack1996} predicts a mass-metallicity relationship, which has been observed in the solar system \cite{atreya2016} and also reported for exoplanets \citep[][]{welbanks_19}. The relatively water poor atmosphere of V1298 Tau b that we find in this work indicates that this planet must have spent most of its accretion phase within the water ice line, thereby failing to accrete volatile rich pebbles \cite{oberg_2011}. The volatile content of the inner disk can be strongly affected by the growth of massive planets in the outer part of the disk \cite{bitsch_21}. In this scenario, a massive planet, formed beyond the water ice line, blocks the supply of volatile rich pebbles in the inner part of the disk, thereby making the inner disk dry and the timescale for core assembly by pebble accretion longer. RV constraints on the mass of V1298 Tau e puts it in a Jupiter/sub-Jupiter range \cite[][]{mascareno_2021,sikora2023,Feinstein2022} with possible orbital period greater than 40 days. Therefore, pebble filtering could play an important role in this system by producing volatile poor atmospheres of the inner planets. 

Alternatively, V1298 Tau b could have accreted volatile rich material that ended being locked up in the interior of the planet. In addition, young planets could experience extreme mass loss driven by intense XUV flux of their active host stars. Using updated mass constraints for this planet we simulated the mass and radius evolution (See Figure \ref{fig:fig3} and Methods Atmospheric Evolution Models). We estimate the Jean's escape parameter \citep{fossati2017}) to be 27. Our calculations suggest that V1298 Tau b is susceptible to photoevaporation in contrast to the conclusions obtained in \citep{maggio2022} based on RV mass estimates. V1298 Tau b may lose up to a few Earth masses within first 1Gyr of its life . Rocky pebble / planetesimal accretion theory of planet formation \cite{ormel2021} predict a gradually mixed interior structure as observed for Jupiter \cite{wahl2017}. We show two possible interior and evolution models for V1298 Tau b (Extended Data Figure \ref{fig:ev}); the core-envelope structure and the diluted core structure. The observed mass, radius and metal poor envelope can all be explained by both models, however in the diluted core scenario, the atmospheric metallicity is expected to evolve due to the removal of the upper layer of the atmosphere due to mass loss as well as convective mixing in the interior which could ultimately reconcile V1298 Tau b with the mature exoplanet population\cite{vazan2022}.

Self-consistent atmospheric modelling for V1298 Tau b shows that we require extremely high internal temperature ($\sim$400K) and strong vertical mixing to explain the non-detection of methane. We show the internal temperature from the early evolution models (Extended Data Figures \ref{fig:ev}) are consistent with theoretical expectations \cite{fortney_2020}. Internal temperatures as high as 300-400K may require external heating mechanisms, such as tidal heating \cite{millholand_2020}. Alternatively, photolytic destruction of methane could also potentially produce a methane poor atmosphere \cite{hu2021}, which may be feasible given the youth and high activity levels of V1298 Tau. We test this hypothesis by running self-consistent forward model using published UV spectrum of V1298 Tau \citep{maggio2023}. However, for V1298 Tau b photochemistry does not impact the methane abundance for pressures higher than $10^{-4}$ bar even for extreme case (1000$\times$solar XUV flux, Supplementary Figure 4).

In conclusion, a strong detection of a water vapour absorption feature in the NIR spectrum of V1298 Tau b allows us to put a stringent upper limit on its mass (24$\pm$5$M_{\oplus}$). The observed spectrum does not exhibit methane feature and is best interpreted with a solar/sub-solar metallicity atmosphere. Leveraging the absence of spectral signature of methane we provide constraints on the internal temperature of the planet. We find that the V1298 Tau system is likely to have formed either late, within the water ice line in a gas-poor, dry and depleting protoplanetary disk, or early in the inner region of the disk with an accretion rate likely moderated by disk gas replenishment or enhanced envelope opacity. V1298 Tau b is likely to undergo atmospheric mass loss and could end up as a Neptune or a low density sub-Neptune or even potentially a super-Earth (see Figure \ref{fig:fig3}).

\section*{Methods} \label{section: methods}

\subsection*{Observation} \label{subsection: observation}

The observations were taken using HST/WFC3 G141 grism in bi-directional spatial scanning mode, covering a range of 1.1-1.7 $\mu$m, with a scan rate of 0.23"/sec. This resulted in 180 exposures over 10 HST orbits. The individual pixels reached a maximum flux level of 30,000 electrons which is roughly 40\% of the saturation level and well within the linear response regime of the detector. We used the $256\times256$ pixel subarray and  \texttt{SPARS25}, \texttt{NSAMP}=5 readout mode which resulted in 88.4 s exposures.

\subsection*{Data reduction} \label{subsection:data reduction}

We use a custom data reduction pipeline for our data analysis \cite{Arcangeli2018,jacobs2022}. The WFC IR detectors are read multiple times non-destructively (without flushing out the accumulated charge) during an exposure. First, sub-exposures are formed for each exposure by subtracting consecutive non-destructive reads and each sub-exposure is reduced separately for improved background subtraction and cosmic ray rejection. We calculate a wavelength solution by matching the first exposure of the visit to a convolution of a PHOENIX stellar spectrum \cite{husser13} for V1298 Tau ($T_{eff}$=4920K) with the response function of G141.

We apply a wavelength dependent flat-field correction and flag bad pixels with data quality DQ=4, 32 or 512 by \texttt{calwf3} and apply a local median filter to identify cosmic rays and clip pixels that deviate more than five median deviations. On average we find 0.53\% pixels affected by cosmic rays for each sub-exposure. To account for the dispersion direction drift of the spectrum we use the first exposure of a visit as a template and shift the spectrum for each exposure along the dispersion direction to match the template. The maximum shift that we measure is 0.3 pixels. Finally, we apply an optimal extraction algorithm \cite{horne1986} on each sub-exposure to maximize signal-to-noise ratio. We shift and shrink the spectra of each sub-exposure to match the wavelength grid of the first sub-exposure by a maximum of 1.05 pixels and 0.65\%.  

\subsection*{Light curve analysis} \label{subsection:light curves}

\begin{figure}[!htb]
    \centering
    \includegraphics[width=0.75\linewidth]{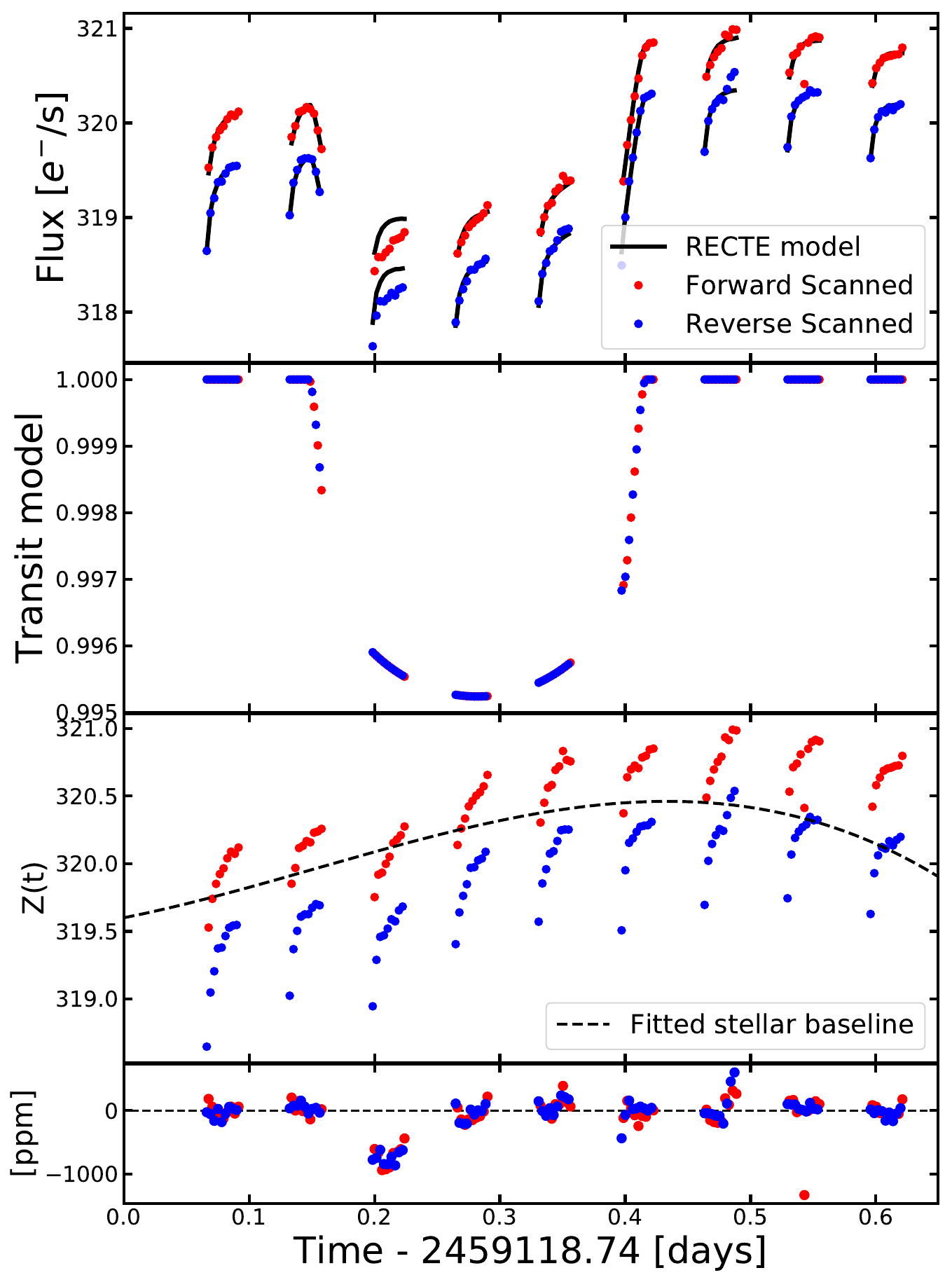}
    \caption{Upper panel: Observed white light curve ($1.1-1.65 \mu m$) of a primary transit of V1298 Tau b and best fit model (black solid lines) Second panel: Best fit planetary transit light curve model. Third panel: Systematics function model estimated by dividing the observed light curve in the upper panel by the best fit transit model shown in the second panel, following the prescription of \cite{kreidberg14}. Black dashed line shows the best fit baseline model. Lower panel: Residual from the white light curve fits. The residuals in the third orbit indicate a possible bright spot crossing. The residuals at the end of the seventh orbit rise sharply. This could potentially be due to a flare event. Red and blue points denote forward and reverse scanned exposures respectively.  Assuming wavelength independent instrumental and stellar systematics, Z(t) is used to de-trend the spectroscopically binned light curves. For further details, See Methods Light curve analysis). }
    \label{fig:white_lc}
\end{figure}

WFC3 light curves are known to exhibit strong time dependent ramp-like (charge-trapping) and visit-long systematics \cite{kreidberg14,Tsiaras_2016a,deming13}. It has been known that the first orbit of each visit has stronger systematics compared to the rest of the visit. Following common practice \citep{wakeford2016}, we exclude this orbit from the rest of the analysis. We modelled the white light curve instrumental systematics using a charge-trapping model, \texttt{RECTE} \citep{Zhou_2017}. The out of transit baseline is a combination of instrumental visit long slopes, well known for HST/WFC3 time series observations \citep{wakeford2016} and rotational variability from the active young host star. Visit-long slopes have been modelled using linear functions in time \citep{wakeford2016,fraine2014}, however the significant non-linearity exhibited by the baseline highlights the effects of stellar variability. We test polynomial functions of first order to fourth order as well as sinusoidal function to model the baseline. A polynomial of third order provides the best fit (lowest BIC value) to the observations. Therefore, model the stellar baseline using a third order polynomial and the stellar disk using a linear limb darkening model. The best fit polynomial function is shown in Extended Data Figure \ref{fig:white_lc} and shows $\sim0.3\%$ variability during the entire visit.  The planetary transit signal is modelled using \texttt{batman} \cite{batman}, where we fix the orbital parameters to known literature values \cite{david19,Feinstein2022,sikora2023}. We ran an MCMC using \texttt{emcee} \cite{emcee} to estimate model parameter uncertainties (Figure \ref{fig:white_lc_corner}.  We find the ninth exposure of the seventh orbit to be affected by a satellite crossing event and exclude this exposure \cite{fu2021}.

\begin{figure}[!htb]
    \centering
    \includegraphics[width=0.75\linewidth]{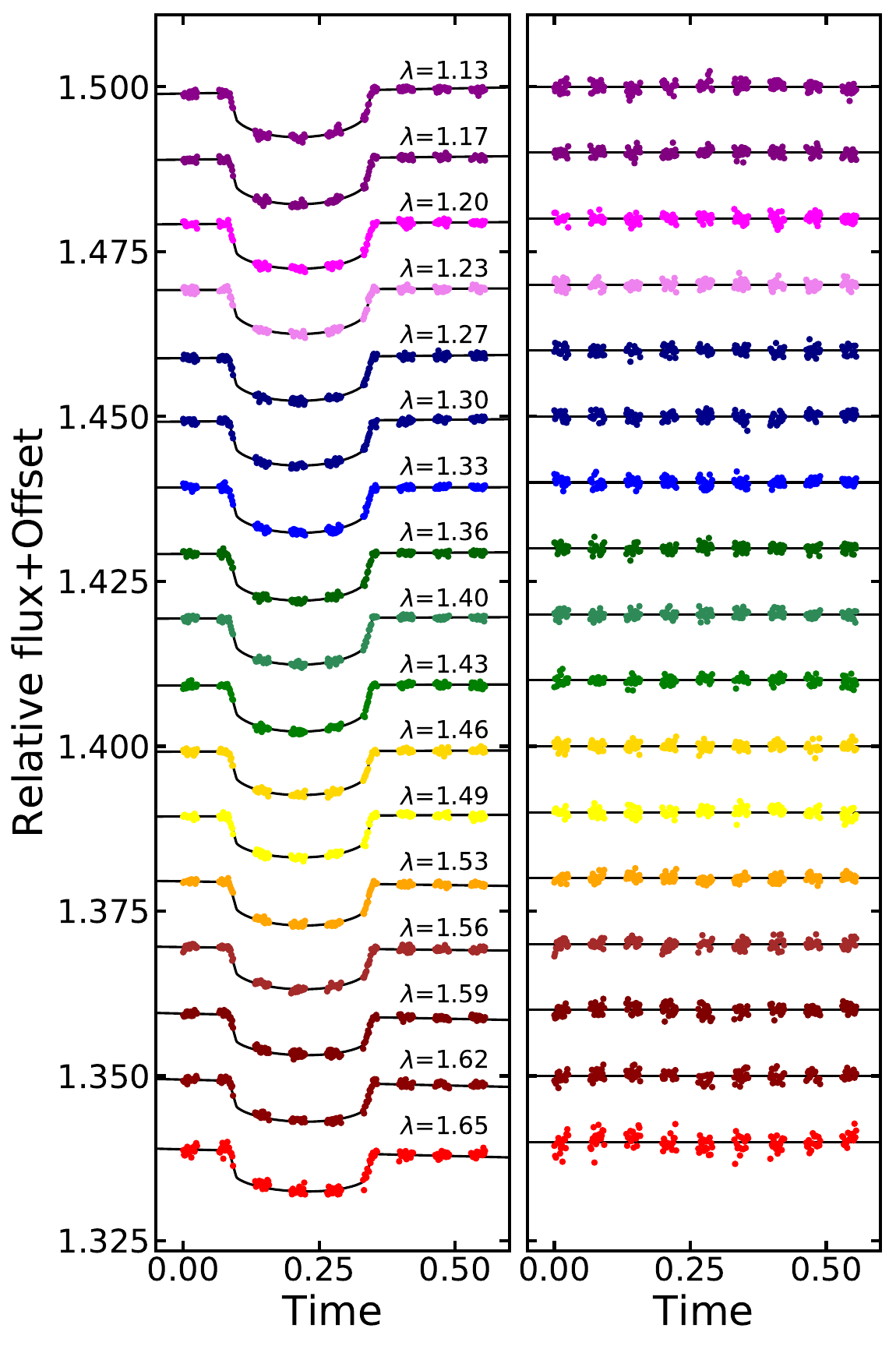}
    \caption{ Left panel: Systematics de-trended spectroscopic light curves along with the best fit transit model. The light curves have been offset vertically for visual clarity. Right panel: Residuals from the light curve fits in the left panel have been shown. For visual clarity, the residuals have been multiplied by three. The residuals are less than 1.3 times the expected photon noise in all spectroscopic channels (except the last one). The systematics de-trending prescription has been described in Methods Light curve analysis. }
    \label{fig:spectroscopic_lc}
\end{figure}

We generate 7 pixel bin spectroscopic light curves from the reduced 1D stellar spectra across 17 wavelength channels. We de-trend the spectroscopic light curves using a common-mode approach given the deviations from the standard HST instrument systematics (possibly due to stellar activity). The common-mode \texttt{divide-white} has been used previously for WFC3 analysis \cite{kreidberg14}; it adopts an agnostic approach to the exact mathematical form of the instrument systematics assuming it is wavelength independent. We model the spectroscopic light curves using a \texttt{batman} model and a linear stellar baseline. We fit for the linear limb darkening coefficient. The observed white light curve, best fit transit model and the derived systematics function are shown in Figure \ref{fig:white_lc}. The systematics de-trended spectroscopic light curves along with the residuals are shown in Figure \ref{fig:spectroscopic_lc}. We also derive the transmission spectrum by fitting each spectroscopic light curve using a \texttt{RECTE} and polynomial stellar baseline models and the derived spectrum agree within 1$\sigma$ to the common-mode spectrum. However, the quality of the fits in the common-mode approach are superior. The residual noise in all the spectroscopic channels is less than 1.3 times the expected photon noise and the average precision on the extracted transit depths is 47ppm.  The fitted transit depths and linear limb-darkening coefficients are shown in Table \ref{tab:table2}. The rms noise is relatively high \cite{stevenson2019}, however this could be a combination of stellar variability, spot crossings and high measured x-shifts.

We note a possible bright spot occultation in the third orbit and also a potential flaring event affecting the latter half of the seventh orbit (Extended Data Figure \ref{fig:white_lc}). To estimate the effect of these exposures on the derived transmission spectrum, we fit the spectroscopic light curves with and without these exposures. We do not find any change in the derived transmission spectrum and the average residuals decrease by 3 ppm when these exposures are excluded.  We conclude that the removal of these exposures do not have a significant manifestation on the spectrum. We also test the effect of the large horizontal drift of the telescope. We incorporate a linear function of x-shifts as a correction factor for the white light curve fits, following the approach of \cite{Tsiaras_2016b}. We find $\Delta$BIC=3 when we include horizontal drift into the fitting algorithm and hence we conclude that including the effect of horizontal drifts is not statistically significant.

\subsection*{Accounting for stellar activity} \label{subsection:stellar activity}

 V1298 Tau is a young pre-main sequence star, known to exhibit 2\% variability in \textit{Kepler} and \textit{TESS} light curves \cite{david19,Feinstein2022,sikora2023}. Variability in such young stars can be attributed photospheric inhomogeneity (star spots and faculae) and fast stellar rotation. Unocculted star spots can contaminate the observed transmission spectrum \cite{barclay_2021}. We estimate the effect of stellar contamination on the transmission spectrum of V1298 Tau b following the prescription of \cite{rackham_2019}. We adopt a surface inhomogeneity model (20\% spot coverage) for V1298 Tau from \cite{feinstein_2021}.  Photospheric temperature contrasts have been studied for T Tauri stars \cite{koen_16}; stars with photospheric temperatures similar to V1298 Tau can have spot temperature contrast upto 1000K. We estimated an extreme case contamination spectrum for V1298 Tau assuming 20\% spot coverage and 1000K spot temperature contrast. The contamination corrected spectrum is consistent within 1$\sigma$ of the uncorrected spectrum. A comparison between the corrected and uncorrected spectra is shown in Extended Data Figures \ref{fig:edwards_comparison}. We re-run retrievals on the contamination corrected transmission spectrum. The retreivals are identical in setup to the uncorrected case (See Methods Atmospheric Models). The posterior distributions of the parameters are shown in Extended Data Figure \ref{fig:retrieval_posterior} with the posteriors from the uncorrected spectrum. All the parameters agree for both cases within 1$\sigma$. The retrieved atmospheric metallicity in the corrected case prefers more sub-solar values compared to the uncorrected case, thereby confirming the robustness of the conclusions drawn in this work. The retrieved paramters are shown in Table \ref{tab:table3}

\begin{figure*}[!htb]
    \centering
    \includegraphics[width=0.65\linewidth]{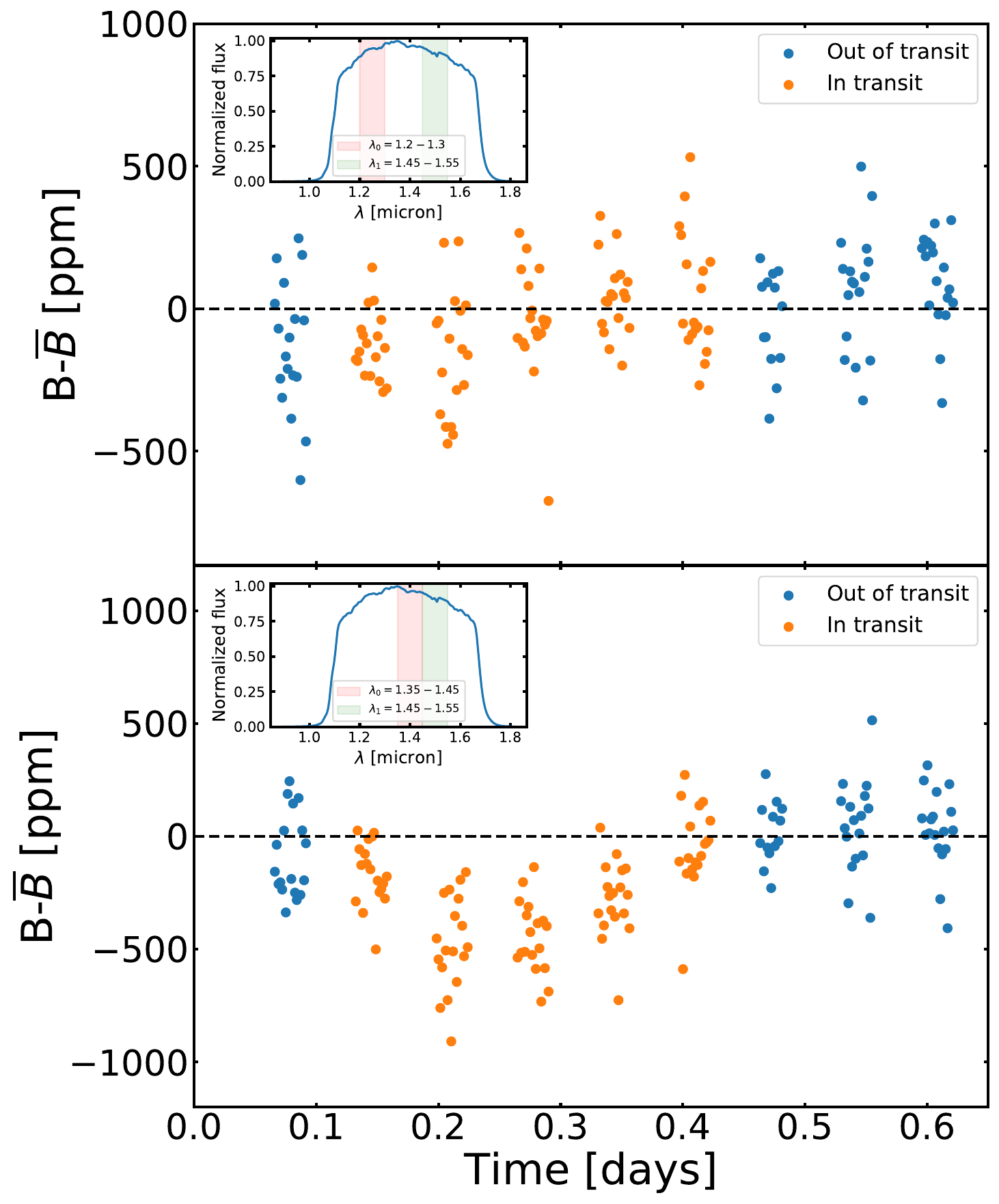}
    \caption{ Excess absorption in the water band during the primary transit of V1298 Tau b. The upper panel shows B-$\overline{B}$ (See Methods, Stellar activity and Equation \ref{eq:eq1}) for 1.25-1.35 $\mu$m and 1.45-1.55 $\mu$m (left and right edge of water absorption band) wavelength bands and the lower panel shows B for the wavelength bands 1.35-1.45 $\mu$m and 1.45-1.55 $\mu$m. The inset panels show the observed 1D stellar spectrum of V1298 Tau b with the bands used for calculating B.  There is an excess absorption in the water band (1.35-1.45$\mu$m) during the in-transit orbits compared to the out of transit orbits, showing that the water absorption is of planetary origin.}
    \label{fig:blueness_criterai}
\end{figure*}

\begin{figure}[!htb]
%\begin{subfigure}
%  \centering
%  \includegraphics[width=0.5\linewidth]{images/RLt_v1298b.png}
%\end{subfigure}
%\begin{subfigure}
  \centering
  \includegraphics[width=\columnwidth]{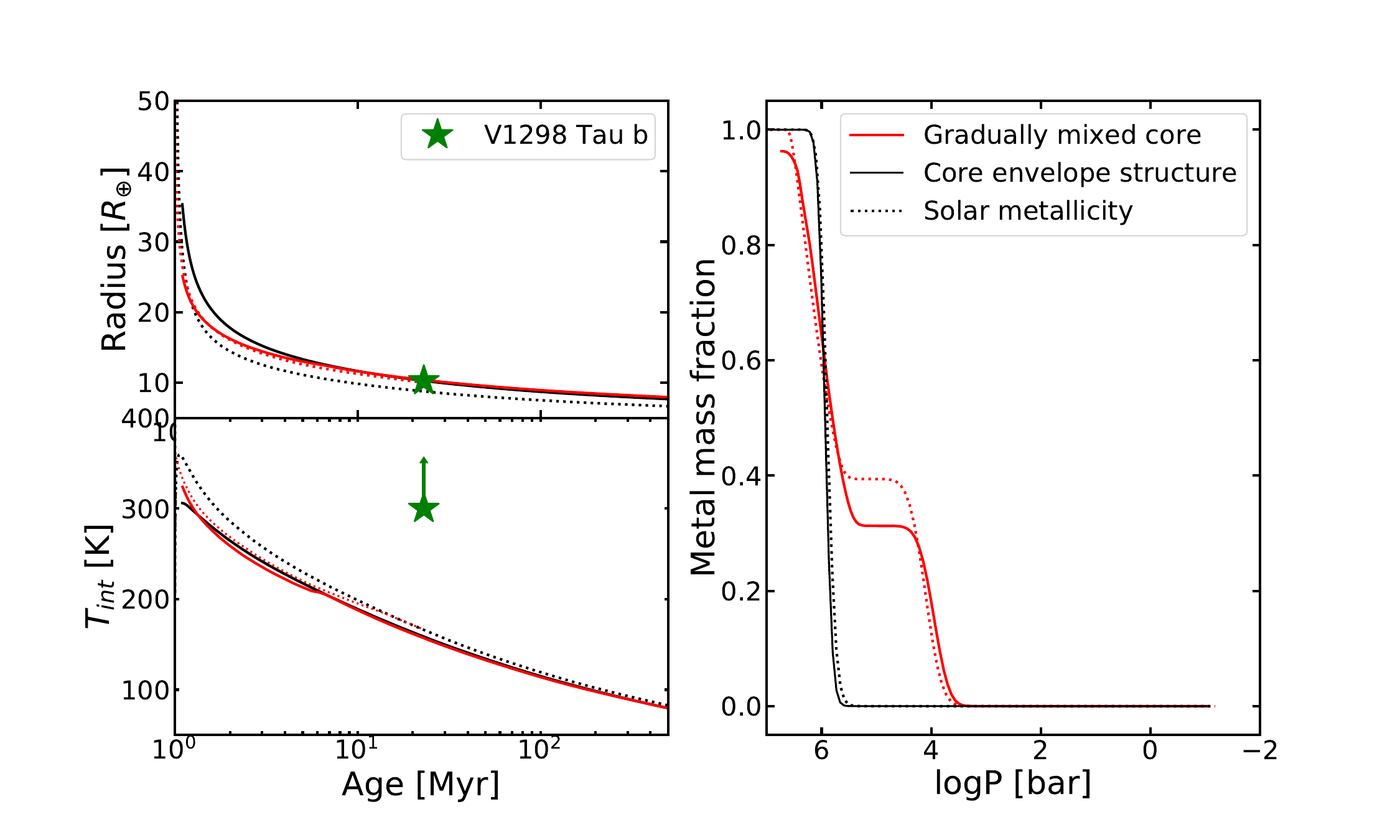}

\caption{Left: Evolution of radius (upper panel) and internal temperature (lower panel) of two possible formation-evolution tracks; core-envelope structure (black line) and diluted core structure (red line) of V1298 Tau b. Solid and dotted lines represent simulations with sub-solar (0.1 solar) and solar envelope metallicity. Right: the metal distribution in the interior as a function of pressure for the two models at its current age (23 Myr). Models are calculated for in-situ formation {of planets with 35-45\% H/He (in mass),} starting from Hill sphere radius. Evolution model is based on \cite{vazan2022}. Both core-envelope models and diluted core models can explain the current size, mass and low metallicity envelope of V1298 Tau.  (See Discussion for more details).}
    \label{fig:ev}
\end{figure}

The effect of stellar absorption has been seen in the limb darkening coefficients \citep[e.g. see][]{kreidberg14}. We set the limb darkening coefficients as a free parameter while fitting for the spectroscopic light curves and the results have been tabulated in Table \ref{tab:table2}. The limb darkening coefficients do not show any effect of stellar absorption. To further convince ourselves that the water absorption feature we see in the spectrum of V1298 Tau b around 1.4$\mu$m is of planetary origin, we define a quantity B as the ratio of flux observed in two wavelength bands.

\begin{equation} \label{eq:eq1}
    B=\frac{\int_{{\lambda}_{1}}^{{\lambda}_{2} }F(\lambda) \,d{\lambda}}{\int_{{\lambda}_{3}}^{{\lambda}_{4} }F(\lambda) \,d{\lambda}}
\end{equation}
 where F is the electron per unit wavelength in the 1D extracted spectra of our reduced exposures, ${\lambda}_{1}$ and ${\lambda}_{2}$ give us lower and upper limit of the first wavelength band and ${\lambda}_{3}$ and ${\lambda}_{4}$ give us lower and upper limit of the second wavelength band.  We calculate B for all the exposures, first using the wavelengths 1.25-1.35 $\mu$m (left end of the water feature) and 1.45-1.55 $\mu$m (right end of water feature) (Upper panel in Figure \ref{fig:blueness_criterai}) and subsequently using 1.35-1.45 $\mu$m (centre of water feature) and 1.45-1.55 $\mu$m (Lower panel Figure \ref{fig:blueness_criterai}).  For the latter case we find an excess absorption during the transit of the planet which indicates that the water absorption of planetary origin.

\subsection*{Atmospheric Models} \label{subsection:atmospheric models}

We use the publicly available 1D radiative transfer code \texttt{PetitRadtrans} to retrieve the atmospheric properties of V1298 Tau b from its observed transmission spectrum. The transmission spectrum does not show methane absorption signature around 1.6$\mu$m which would be expected for a warm planet like V1298 Tau b based on equilibrium chemistry. The lack of methane can be explained by disequilibrium processes, like vertical mixing \cite{fortney_2020,Baxter2021} dredging up methane poor gas from the hot interior parts of the atmosphere. In our retrieval framework, we modelled this effect using a `quenching pressure', where VMR of C, H, O, N bearing molecules are calculated using \texttt{PetitRADTRANS}., however above the quench point, the molecular concentrations are held constant. We model the atmospheric thermal structure with a Guillot T-P profile \cite[][]{guillot2010} shown in Eqn \ref{eq:eq2}, where $T_{equ}$ and $T_{int}$ are the equilibrium and internal temperature of the planet. ${\kappa}_{IR}$ is the average infrared atmospheric opacity and $\gamma$ is the ratio between optical and IR opacity.  We constrain the models by fixing the the values of both ${\kappa}_{IR}$ to 0.01 $cm^{2} g^{-1}$ and $\gamma$ to 0.01, assuming the atmospheric opacity at the observed band pass to be water dominated. We include $H_{2}O$, C$H_{4}$, C$O_{2}$ and CO opacities in our retrieval framework as these molecular species have absorption features in the NIR \citep{madhusudhan2019}. We do not include HCN, N$H_{3}$ opacities in our retrievals as we do not find evidence of these species in free retreivals (See Supplementary Figure 1). We assume a grey cloud deck opacity model to simulate cloud absorption.
\begin{equation} \label{eq:eq2}
    T^{4}=\frac{3T_{int}^{4}}{4} \left( \frac{2}{3}+\tau \right) + \frac{3T_{equ}^{4}}{4} \left [\frac{2}{3}+\frac{1}{\gamma \sqrt{3}}+ \left(\frac{\gamma}{\sqrt{3}}-\frac{1}{\gamma \sqrt{3}}\right)e^{-\gamma \tau \sqrt{3}} \right], \nonumber
\end{equation} 
 \begin{equation} \label{eq:eq3}
     \tau = P \kappa_{IR}/g
 \end{equation}

We fix the mass of the planet to 24$M_{\oplus}$, based on the mass upper limit estimated from the scale height.  The free parameters in our models are atmospheric metallicity, C/O ratio, $R_{p}$ (radius of planet at reference pressure, 1 bar), $T_{equ}, T_{int}$, $P_{quench}$ and $P_{cloud}$. We run an MCMC with 3,000 burn in steps and 30,000 post burn-in steps with 50 walkers. We put uniform priors on the fitting parameters. The posterior distribution of the fitted parameters is shown in Figure \ref{fig:retrieval_posterior}. We retrieve a sub-solar/solar metallicity. The retrieved equilibrium temperature is consistent with the expected equilibrium temperature of the planet. The retrieved parameters have been summarized in Table \ref{tab:table3}. 

We test the importance of the internal temperature by fixing the internal temperature to 0K (i.e fitting for an isothermal atmosphere). High internal temperature models are statistically favoured by a $\Delta$BIC=50.  We also perform free retrievals using an isothermal atmosphere (See Supplementary Information Figure 1). This yields an upper limit to the methane Volume Mixing Ratio (VMR) in the atmosphere ($\sim 10^{-6}$) which is lower than the detection threshold for HST \citep{fortney_2020}, thereby independently confirming the non-detection of methane. The free retrieval did not find evidence for other molecular species like HCN and N$H_{3}$ putting upper limits of $10^{-6}$ on their VMRs.   We explore the effect of fixing the planet's mass in Extended Data Figure \ref{fig:mass_uncertainty} and Methods Mass estimate.

 We construct self-consistent atmospheric models with varying internal temperatures to study the quenching of methane and CO in the deep atmosphere. We compute the T-P profile using \texttt{petitCODE} \citep{MolliereEtal2015apjModelatmospheres, MolliereEtal2017aaObservingJWST}, assuming radiative-convective equilibrium. Irradiation onto the planet is computed assuming a planetary-wide energy redistribution, with a host star effective temperature and radius of 4970~K and 1.31~R$_\odot$, semi-major axis of 0.1688~AU, and planetary intrinsic temperatures of 100~K -- 400~K. Using our retrievals as a guidance, a solar metallicity was adopted with a slightly sub-solar C/O of 0.3. We achieved this C/O by reducing the carbon abundance from its solar value. The resulting temperature profiles are shown in Supplementary Figure~\ref{fig:temp_tint_varied}. Subsequently, we use a 1D chemical kinetics model \citep{AgundezEtal2014aaPseudo2D} in combination with a photochemical network \citep{VenotEtal2022aaScheme} to calculate self-consistent vertical quenching pressures for the main atmospheric species. We perform our calculations with a constant eddy diffusion coefficient ($K_{zz}$) of $10^{10}$~cm$^2$/s. This value, although high, is in line with the expected values for convective mixing in giant planets and brown dwarfs \citep[e.g.][]{FreytagEtal2010aaConvection, MosesEtal2011apjDisequilibrium, Zhang2020raaAtmosphericRegimesReview}. We include photochemistry in our models, however, we find that it does not significantly affect the molecular abundances at pressures typically probed by  transmission spectroscopy. We test the effect of higher XUV flux of host star by computing models for scaled solar spectra (10-1000 times, Supplementary Figure 3,4). The resulting chemical disequilibrium abundances for methane, CO, and water are shown in Supplementary Figure~\ref{fig:chem_tint_varied}. We find that the planet should have high internal temperature ($\sim$ 300-400K) to have the carbon chemistry to be CO dominated. This is consistent with the high internal temperature and deep quenching concluded from the retrieval analysis. 

\subsection*{Mass estimate} \label{subsection:mass estimate}

We estimate the mass of V1298 Tau b from the transmission spectrum using the approach described in \cite{de_wit}. 

\begin{equation} \label{eq:eq4}
    M_{p}=\frac{kT{R_{p}^{2}}}{{\mu}GH}
\end{equation}

We use the radius measurement from \textit{Kepler} \citep[][]{david_19_b} and an equilibrium temperature of 670K for the calculation. We estimate the scale height from the observed spectrum of V1298 Tau b. The height of an atmosphere can be estimated using Equation 1 of \cite[][]{etangs08}:

\begin{equation} \label{eq:eq5}
    z({\lambda}_{2})-z({\lambda}_{1})=Hln \left(\frac{\sigma({\lambda}_{2})}{\sigma({\lambda}_{1})}\right)
\end{equation}

In Eqn \ref{eq:eq5}, $\sigma$ is the absorption cross section at a given wavelength, z is the measured radius of the planet at a given wavelength. We estimate 2.7 scale heights to account for the 1.4$\mu$m water absorption feature, assuming a water dominated atmospheric opacity and a cloud free atmosphere. Given the young age and inflated size, we assume a primordial H/He rich atmosphere and fix the mean molecular mass to 2.33. We find a large atmospheric scale height for V1298 Tau b; 1000 $\pm$ 200 km and a mass estimate of 24$\pm$5$M_{\oplus}$. The reported radius of V1298 Tau b differs slightly between measurements from different epochs. K2 \citep{david_19_b}
 and TESS \citep{Feinstein2022} differ by $\sim 2\sigma$, whereas \citep{sikora2023}
 estimate the radius to be in between. We estimated the planet mass using all the three radius measurements, the results of which are tabulated in Supplementary Information Table 1. We also used the HST white light curve radius (0.84$\pm$0.003$R_{J}$ to estimate the planet. All the estimates are consistent with each other within 1$\sigma$, and to be conservative we adopt the highest estimate of K2 (24$\pm$5$M_{\oplus}$).
 
This estimate can be interpreted as an upper limit, given the assumption of cloud free. In case of a cloudy atmosphere, the measured scale height from the spectrum would be underestimated, therefore 
 leading to an over estimation of the mass. Given the observed spectrum, a cloud-free case would therefore yield the maximum possible mass for this planet. 
 
\begin{figure}
    \centering
   \includegraphics[width=\columnwidth]{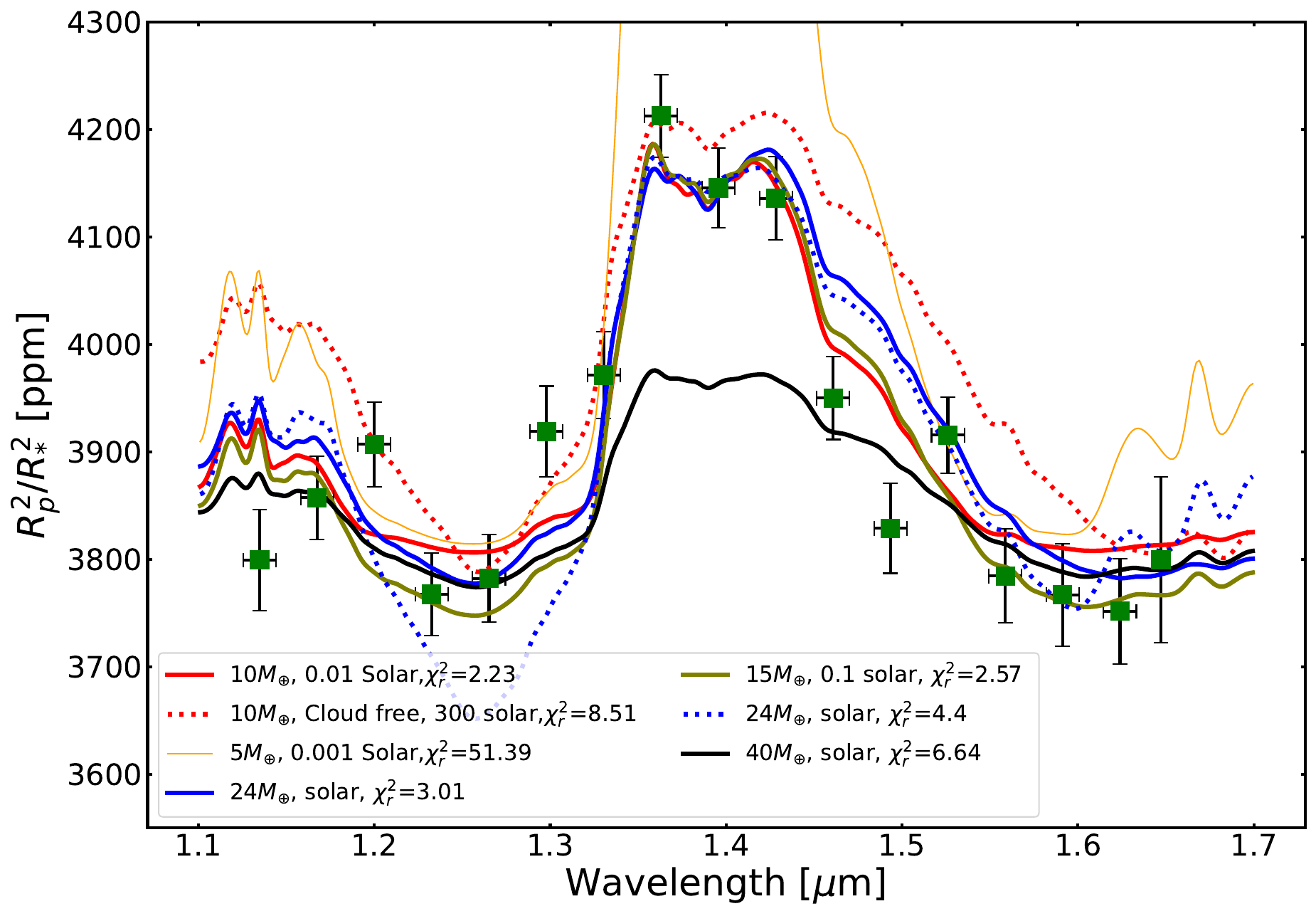}
    \caption{ Comparison between the observed transmission spectrum of V1298 Tau b and retrievals using \texttt{PetitRADTRANS} \cite{molliere_2020} with different planet masses. The red, green and blue models represent retrieved median models for 10, 15 and 24$M_{\oplus}$ models with a grey cloud opacity respectively. The corresponding dotted line show cloud free models at the same mass. The orange model represents a 5$M_{\oplus}$ model; our retrievals failed to converge for this case. The black solid line shows a model with 40$M_{\oplus}$. The observation can be fitted with 24, 15 and 10$M_{\oplus}$ models. For the lower mass cases our retrievals converged on extremely low atmospheric abundances (0.1-0.01 solar) to fit the water absorption feature. Our retrievals statistically favour cloudy models.  }
    \label{fig:mass_uncertainty}
\end{figure}

\begin{figure}
    \centering
    \resizebox{12cm}{!}{\includegraphics[width=\linewidth]{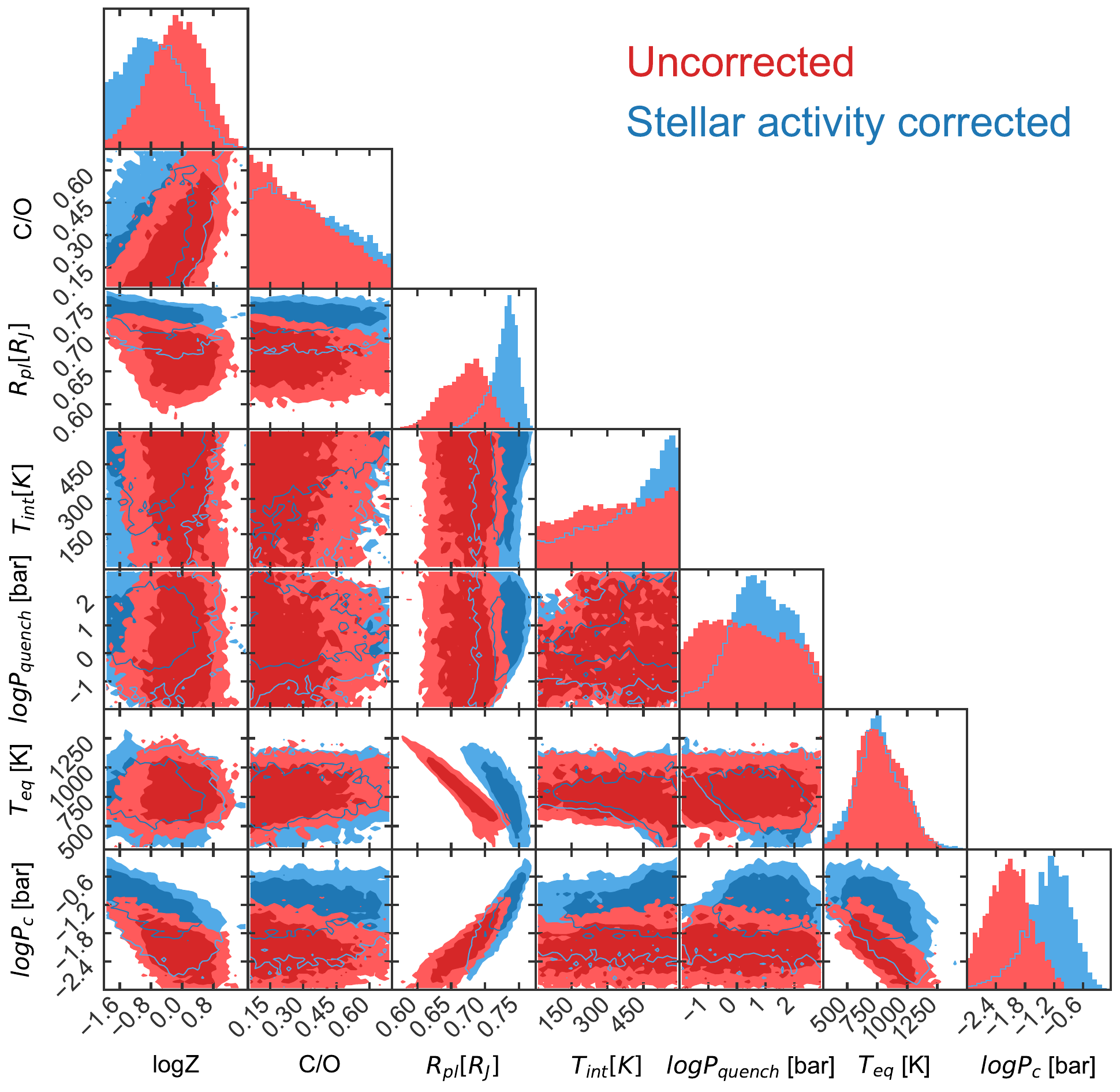}}
    \caption{The posterior distribution from retrieval done on the uncorrected (red) and stellar activity corrected (blue) transmission spectrum of V1298 Tau b assuming a mass of 24$M_{\oplus}$. We used 1D atmosphere model,  Guillot T-P profile \cite{guillot2010}, equilibrium chemistry with atmospheric quenching (See Methods Atmospheric models for details). We retrieve an atmospheric metallicity consistent with sub-solar/solar for both cases. The retrieved parameters with their 1$\sigma$ confidence intervals are shown in Table \ref{tab:table3}. The stellar activity corrected spectrum is shown in comparison with the uncorrected spectrum in Extended Data Figure \ref{fig:edwards_comparison}. In the same figure, the contamination function \citep{rackham_2019} used for correcting the spectrum is als shown.  }
    \label{fig:retrieval_posterior}
\end{figure}

To estimate the impact on atmospheric parameters, we run retrievals on the observed transmission spectrum by fixing the temperature to 670K, for different masses (24, 15, 10, 5$M_{\oplus}$) and both cloud free and cloudy cases (Extended Data Figure \ref{fig:mass_uncertainty}). We include the same molecules compared to the 24$M_{\oplus}$ case, as molecular opacities do not depend on the planet's gravity. For the  5$M_{\oplus}$ case, our retrievals did not did converge as it could not reproduce the water absorption signal. We can however fit the observations with 24,15 and 10$M_{\oplus}$ models. We find that cloudy models are statistically favored compared to cloud free models. The 24$M_{\oplus}$ (mass upper limit) model converges at solar atmospheric metallicity; for lower mass models our retrievals converge at even lower (0.1-0.01 solar) metallicities to fit the water absorption feature. We test the robustness of the estimated mass upper limit by running a retrieval with 40$M_{\oplus}$. This model fails to reproduce the observed water feature and can be rejected at high confidence. To further test the robustness of the mass estimate, we run an atmospheric retrieval, keeping mass as a free parameter. The posterior distribution is shown in Supplementary Information  Figure 2. The mass posterior peaks around 10$M_{\oplus}$. The metallicity in this case yields an upper limit of solar value at 2$\sigma$. Therefore, the conclusions of mass less than 24$M_{\oplus}$ and solar/sub-solar atmospheric metallicity appear robust based on this test.

Thus, from the transmission spectrum we can estimate a robust mass upper limit, and conclude that V1298 Tau b is likely to be Neptune or a low-density sub-Neptune or potentially a super-Earth progenitor \citep{poppenhaeger_21}.

\subsection*{Atmospheric evolution models}
 The atmospheric evolution models shown in Figure \ref{fig:fig3}  have been simulated using the open source \texttt{platypos} code \cite{poppenhaeger_21}. The code calculates the mass loss rate at a given point in time, using the energy-limited mass loss formalism \cite[e.g see][]{owen_jackson2012, salz2015} and evolve the planet's physical properties (mass and radius) at every step of the calculation. The radius evolution is a combined effect of atmospheric contraction and mass loss, and the updated size of the planets are calculated from the scaling relation given in \cite{Lopez_2014}. We adopt the stellar luminosity from \cite{poppenhaeger_21}. These simulations have performed considering the estimated mass upper limit (24$M_{\oplus}$). For lower masses we can expect higher mass loss rates.

\subsection*{Comparison with Edwards (2022)}

V1298 Tau b was included in a sample of 70 transiting exoplanets whose spectra have been shown in \cite{edwards2022}. The authors use a different pipeline (\texttt{Iraclis} \cite{Tsiaras_2016b}) for the data reduction. The authors also use a common-mode approach to derive the spectrum of this planet. The transmission spectrum obtained in this work is consistent within 1$\sigma$ to the results of \cite{edwards2022} except a constant offset of $\sim$ 500 ppm. The constant offset is a result of \cite{edwards2022} using the third orbit in their white light curve fits which we choose to exclude because of a potential spot crossing in that orbit. We tested the effect of including the third orbit in the white light curve fits. We find a $\Delta$BIC=170 in favour of excluding the third orbit from the fits. The transmission spectrum obtained by \cite{edwards2022} and this work have been shown together for comparison in Figure \ref{fig:edwards_comparison}. 
%We emphasize that \cite{edwards2022} study focuses on the survey and does not provide a detailed interpretation for the spectrum of V1298 Tau b as done in the current work.

%\subsection{Calculation of internal luminosity}

%The total internal luminosity of a planet is a combination of residual heat of formation and any external sources such as tidal heating. The residual energy of formation is given by \citep[][]{rogers2010}:

%\begin{equation} \label{\eq:eq6}
 %   log \left(\frac{L_{int}}{L_{\odot}}\right)=a_{1}+a_{M}log \left(\frac{M_{p}}{M_{\oplus}}\right) + a_{R}log \left(\frac{R_{P}}{R_{J}}\right) + a_{t}log \left(\frac{t_{p}}{1 Gyr}\right)
%\end{equation}

%where $a_{1}$=-12.46 $\pm$ 0.05, $a_{M}$=1.74 $\pm$ 0.03, $a_{R}$=-0.94 $\pm$ 0.09 and $a_{t}$=-1.04 $\pm$ 0.04 where $t_{p}$ is the age of the planet in Gyr. The tidal luminosity depends on the eccentricity of the planet, tidal quality factor and planetary spin obliquity. The expressions for the calculation of $L_{tide}$ are given in \citep[][]{millholand_2020}. To estimate the internal temperature we assume the total luminosity due to residual energy of formation and tidal heating  to be dissipated due to blackbody emission:

%\begin{equation} \label{eq:eq7}
%    T_{int}=\left(\frac{L_{int}}{4\pi \sigma R_{p}^{2}}\right)^{1/4}
%\end{equation}

\newpage
\section*{Extended Data Figures}

\newpage

\newpage

\newpage

\iffalse
\begin{figure}
    \centering
    \resizebox{12cm}{!}{\includegraphics[width=\linewidth]{images/wavelength_posterior_1.4_.pdf}}
    \caption{Posterior distribution of the parameters used to to fit the spectroscopically binned light curves for the 1.35 micron bin. The light curve was fit using a common-mode method (See Methods section 7.3), and the fitting parameters are transit depth, linear limb darkening coefficient, and the linear stellar baseline. See Methods Section 7.3 for more details. }
    \label{fig:fig9}
\end{figure}
\fi

\begin{figure} 
    \centering
    \resizebox{12cm}{!}{\includegraphics[width=\linewidth]{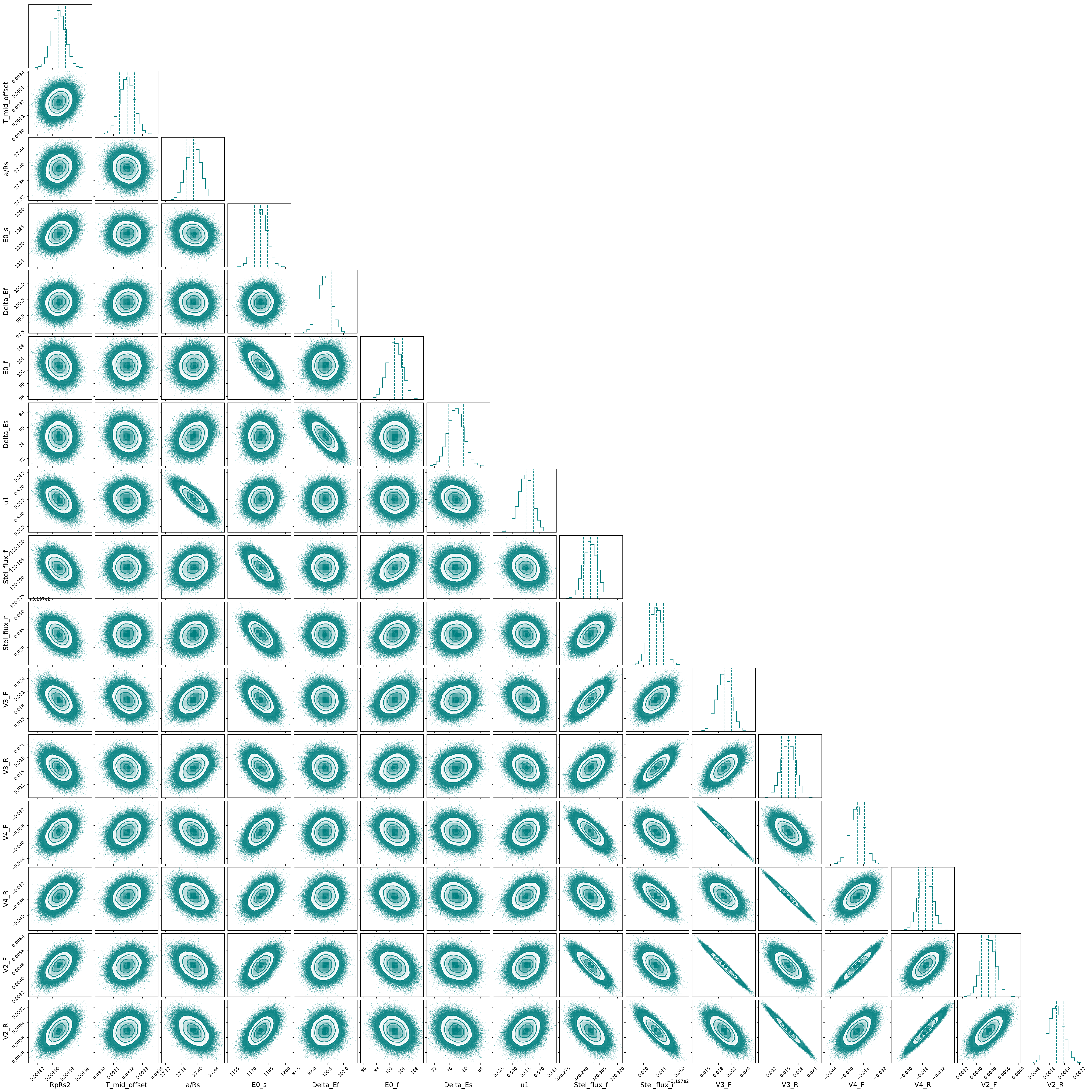}}
    \caption{The posterior distributions of the parameters from the MCMC fits to the white light curves of V1298 Tau b. See Methods Light curve analysis for more details. }
    \label{fig:white_lc_corner}
\end{figure}

\begin{figure}
    \centering
    \resizebox{12cm}{!}{\includegraphics[width=\linewidth]{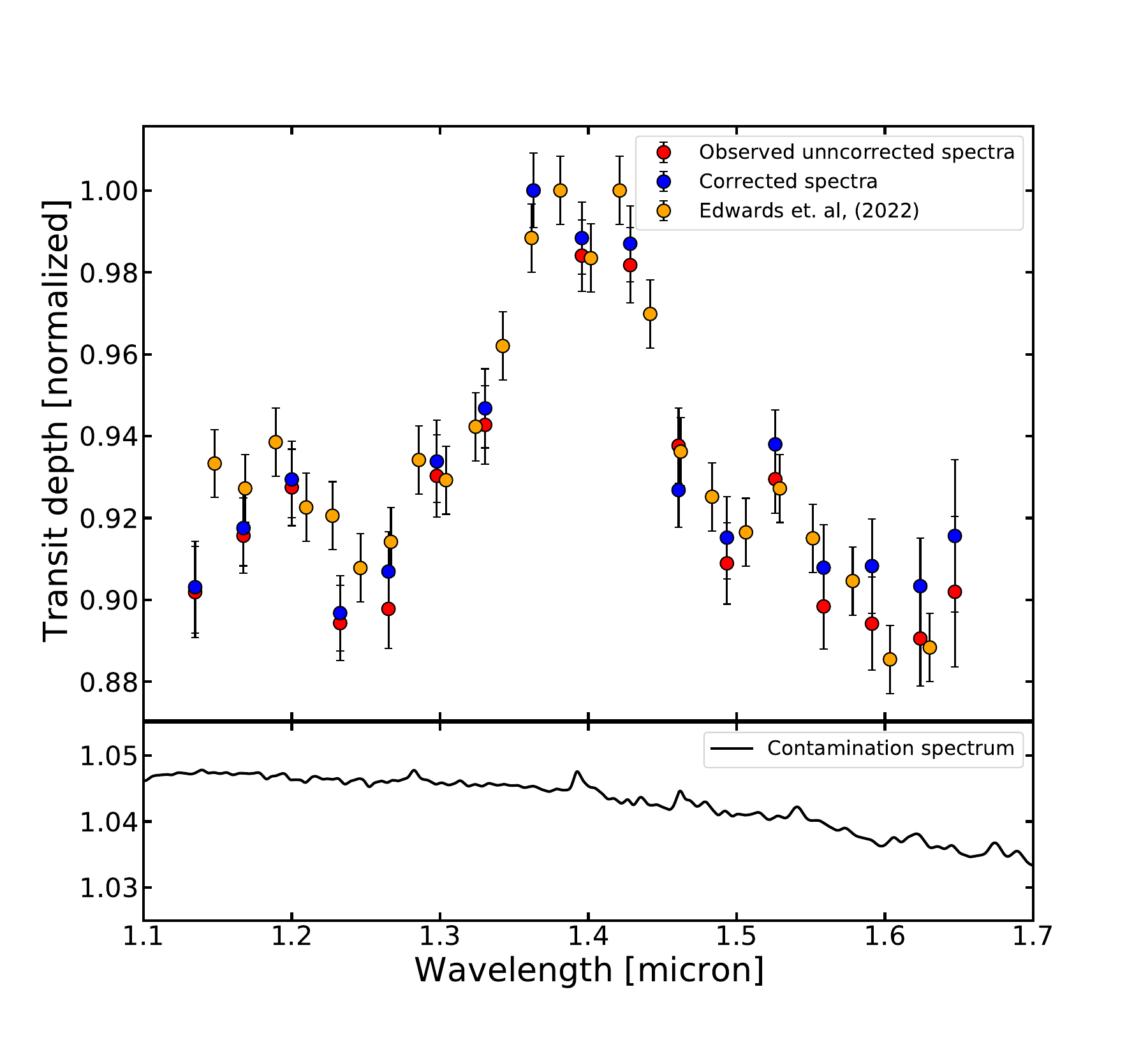}}
    \caption{Upper panel: Comparison of transmission spectrum obtained in this work (red markers) and \cite{edwards2022} (orange markers) for V1298 Tau b. There is a constant offset of $\sim$500ppm between the two spectra, however the atmospheric signature in both are consistent with each other. For aid of visual comparison, both spectra have been normalized. The blue markers show transmission spectrum of V1298 Tau b obtained by applying stellar contamination correction \citep{rackham_2019} on the spectrum derived in this work. The corrected and uncorrected spectra are consistent within 1$\sigma$ for all spectroscopic channels. Bottom panel: Stellar contamination function used to correct the observed transmission spectrum (See Methods Stellar Activity Section).  }
    \label{fig:edwards_comparison}
\end{figure}
\iffalse
\begin{figure}
    \centering
    \resizebox{12cm}{!}{\includegraphics[width=\linewidth]{images/comparison_spectra_stel_correction.pdf}}
    \caption{The upper panel shows the comparison of the uncorrected and stellar activity corrected transmission we have derived for V1298 Tau b. The lower panel shows the contamination function we have used to correct the observed transmission spectrum.  }
    \label{fig:stel_correct_comparison}
\end{figure}

\begin{figure}
    \centering
    \resizebox{12cm}{!}{\includegraphics[width=\linewidth]{images/V1298b_corner_GTC_stel_corrected.pdf}}
    \caption{Posterior distribution of retrievals run on stellar activity corrected transmission spectrum.  }
    \label{fig:stel_corrected_posterior}
\end{figure}

%\iffalse
\begin{figure}
    \centering
    \resizebox{12cm}{!}{\includegraphics[width=\linewidth]{images/V1298b_corner_GTC_edwards_spectra.pdf}}
    \caption{Posterior distribution of retrievals run on transmission spectra obtained with binning similar to \cite{edwards2022} and including the 3rd orbit in the white light curve fitting.  }
    \label{fig:fig14}
\end{figure}

\fi
\begin{figure}
    \centering
    \resizebox{12cm}{!}{\includegraphics[width=\linewidth]{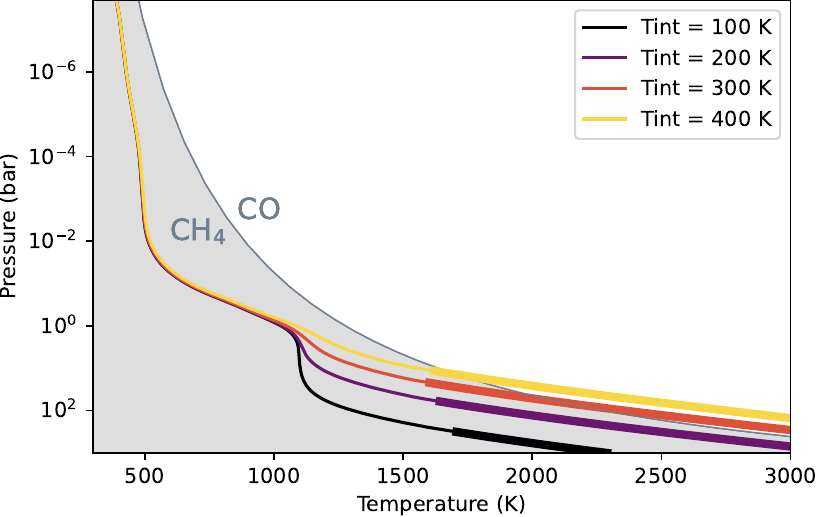}}
    \caption{Pressure-temperature profiles for V1298~Tau~b computed using \texttt{petitCODE}  \cite{molliere2015} for intrinsic temperatures ($T_{int}$) between 100~K and 400~K. The lines are thick in the convective region of the atmosphere. The gray line and background color denote the regions where either methane or CO are the dominant molecule in chemical equilibrium.}
    \label{fig:temp_tint_varied}
\end{figure}

\begin{figure}
    \centering
    \resizebox{12cm}{!}{\includegraphics[width=\linewidth]{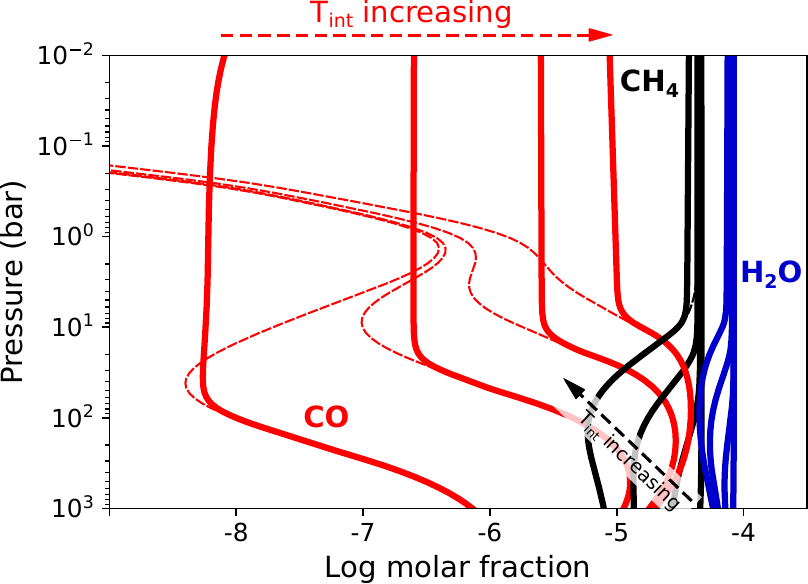}}
    \caption{Chemical abundances of methane (black), CO (red), and water (blue) in the deep atmosphere, for $T_{int} = 100$~K, 200~K, 300~K, and 400~K. Chemical abundances have been calculated using a self-consistent framework with \texttt{petitCODE} and a chemical kinetics model in combination with a photochemical network \cite{AgundezEtal2014aaPseudo2D,VenotEtal2022aaScheme}.Arrows denote increasing $T_{int}$. Molecular abundances are shown for a composition in chemical equilibrium (dashed) and for chemical kinetics, i.e.~when vertical quenching is included (solid). We included photochemistry in our models, however it did impact the molecular abundances at observable pressures.}
    \label{fig:chem_tint_varied}
\end{figure}

%\section{Extended Data tables}

\begin{table}
\caption{Table showing the best fit transit depths, linear limb darkening coefficients and RMS residual compared to expected photon noise for V1298 Tau b. }
\label{tab:table2}
\begin{tabular}{c|c|c|c}
\hline
 Central Wavelength ($\mu$m)  & Transit depth ($R_{p}^{2}/R_{*}^{2}$)  & u1 & Residual (photon noise) \\
\hline
\hline
1.13 & 3766 $\pm$ 49 & 0.59 $\pm$ 0.031 & 1.29 \\

1.16 & 3803 $\pm$ 41 & 0.58 $\pm$ 0.025 & 1.06 \\

1.20 & 3872 $\pm$ 42 & 0.53 $\pm$ 0.026 & 1.08 \\

1.23 & 3757 $\pm$ 40 & 0.58 $\pm$ 0.025 & 1.04 \\

1.26 & 3752 $\pm$ 43 & 0.56 $\pm$ 0.027 & 1.12 \\

1.30 & 3869 $\pm$ 45 & 0.56 $\pm$ 0.028 & 1.16 \\

1.33 & 3945 $\pm$ 43 & 0.55 $\pm$ 0.026 & 1.11 \\

1.36 & 4175 $\pm$ 41 & 0.53 $\pm$ 0.023 & 1.05 \\

1.39 & 4110 $\pm$ 39 & 0.56 $\pm$ 0.023 & 1.01 \\

1.43 & 4111 $\pm$ 41 & 0.57 $\pm$ 0.024 & 1.05 \\

1.46 & 3906 $\pm$ 41 & 0.57 $\pm$ 0.025 & 1.06 \\

1.49 & 3819 $\pm$ 43 & 0.50 $\pm$ 0.028 & 1.13 \\

1.52 & 3849 $\pm$ 40 & 0.57 $\pm$ 0.023 & 1.02 \\

1.55 & 3720 $\pm$ 46 & 0.58 $\pm$ 0.03 & 1.19 \\

1.59 & 3706 $\pm$ 50 & 0.50 $\pm$ 0.034 & 1.30 \\

1.62 & 3682 $\pm$ 51 & 0.54 $\pm$ 0.033 & 1.33 \\

1.64 & 3702 $\pm$ 79 & 0.53 $\pm$ 0.053 & 1.36 \\

\hline
\hline
\end{tabular}
\end{table}

\begin{table}
\centering
\caption{Retrieved atmospheric parameters of V1298 Tau b from its transmission spectrum. See Methods Atmospheric models for details of atmospheric models used.}
\label{tab:table3}
\begin{tabular}{c c c c c}
\hline
Retrieval parameter & 24$M_{\oplus}$ & 15$M_{\oplus}$ & 10$M_{\oplus}$ & 24$M_{\oplus}$ (corrected)
\\
\hline
\hline
logZ ($Z_{\odot}$) & -$0.1^{+0.66}_{-0.72}$ & -$0.9^{+0.9}_{-1.1}$ & -$1.5^{+0.9}_{-1.1}$ & -$0.73^{+0.81}_{-0.76}$ \\

C/O ratio & $0.23^{+0.25}_{-0.13}$ & $0.24^{+0.24}_{-0.13}$ & $0.26^{+0.25}_{-0.15}$ & $0.29^{+0.23}_{-0.17}$ \\

$R_{p}$ [$R_{J}$] & $0.68^{+0.03}_{-0.04}$ & $0.70^{+0.02}_{-0.02}$ & $0.65^{+0.02}_{-0.01}$ & $0.73^{+0.03}_{-0.04}$\\

$T_{int}$ [Kelvin] & $335^{+185}_{-210}$ & $380^{+150}_{-250}$ & $400^{+150}_{-250}$ & $429^{+135}_{-253}$ \\

$logP_{quench}$ [bar] & $0.26^{+1.68}_{-1.36}$ & $0.42^{+1.10}_{-0.63}$ & $0.46^{+1.30}_{-0.78}$ & $0.81^{+1.16}_{-0.11}$ \\

$T_{eq}$ [Kelvin] & $760^{+210}_{-160}$ & 670 (fixed) & 670 (fixed) & $757^{+230}_{-180}$ \\

log$P_{cloud}$ [bar] & $-2.1^{+0.41}_{-0.47}$ & $-1.6^{+0.54}_{-0.57}$ & $-2.0^{+0.5}_{-0.6}$ & $-1.28^{+0.32}_{-0.6}$ \\

\hline
\hline
\end{tabular}
\end{table}
\backmatter

%\bmhead{Supplementary information}

\section*{Data availability}
The data used in this study may be obtained from the Mikulski Archive
for Space Telescopes (MAST, \url{https://mast.stsci.edu/}) and are associated
with HST GO 16083. 

\section*{Code availability}
This research made use of public software like \texttt{astropy}, \texttt{lmfit}, \texttt{emcee}, \texttt{platypos}, \texttt{petitCODE} and \texttt{PetitRADTRANS}

\section*{Correspondence and request for materials}

All correspondences and requests for materials related to this article should be directed to Saugata Barat (Email: s.barat@uva.nl)

\section*{Acknowledgments}

J.M.D acknowledges support from the Amsterdam Academic Alliance (AAA) Program, and the European Research Council (ERC) European Union’s Horizon 2020 research and innovation program (grant agreement no. 679633; Exo-Atmos). This work is part of the research program VIDI New Frontiers in Exoplanetary Climatology with project number 614.001.601, which is (partly) financed by the Dutch Research Council (NWO).
AV acknowledges support by ISF grants 770/21 and 773/21.

%\section*{Declarations}

\section*{Authors' contributions}

SB led the analysis and interpretation of the observations and wrote the majority of the manuscript. JMD designed the project, contributed to the observation proposal and planning, to the analysis, the interpretation and to the manuscript. AV and RB contributed theoretical models of interior structures and atmospheric chemistry respectively, to the analysis, the interpretation and to the manuscript. MRL, JJF, TJD, JHL, JS, HS, EAP, GM contributed to the analysis and manuscript preparation. BJ, VP, LP and KT contributed to the observation proposal and planning and contributed to the manuscript.

.

\noindent

\bigskip
\iffalse
\begin{flushleft}%
Editorial Policies for:

\bigskip\noindent
Springer journals and proceedings: \url{https://www.springer.com/gp/editorial-policies}

\bigskip\noindent
Nature Portfolio journals: \url{https://www.nature.com/nature-research/editorial-policies}

\bigskip\noindent
\textit{Scientific Reports}: \url{https://www.nature.com/srep/journal-policies/editorial-policies}

\bigskip\noindent
BMC journals: \url{https://www.biomedcentral.com/getpublished/editorial-policies}
\end{flushleft}

\fi

\newpage

\vspace{10mm}

\bibliography{sn-bibliography}% common bib file

\newpage
\begin{center}
    \Large{Supplementary Information}

\end{center}

\begin{figure}[H]
    \centering
    \includegraphics[width=\linewidth]{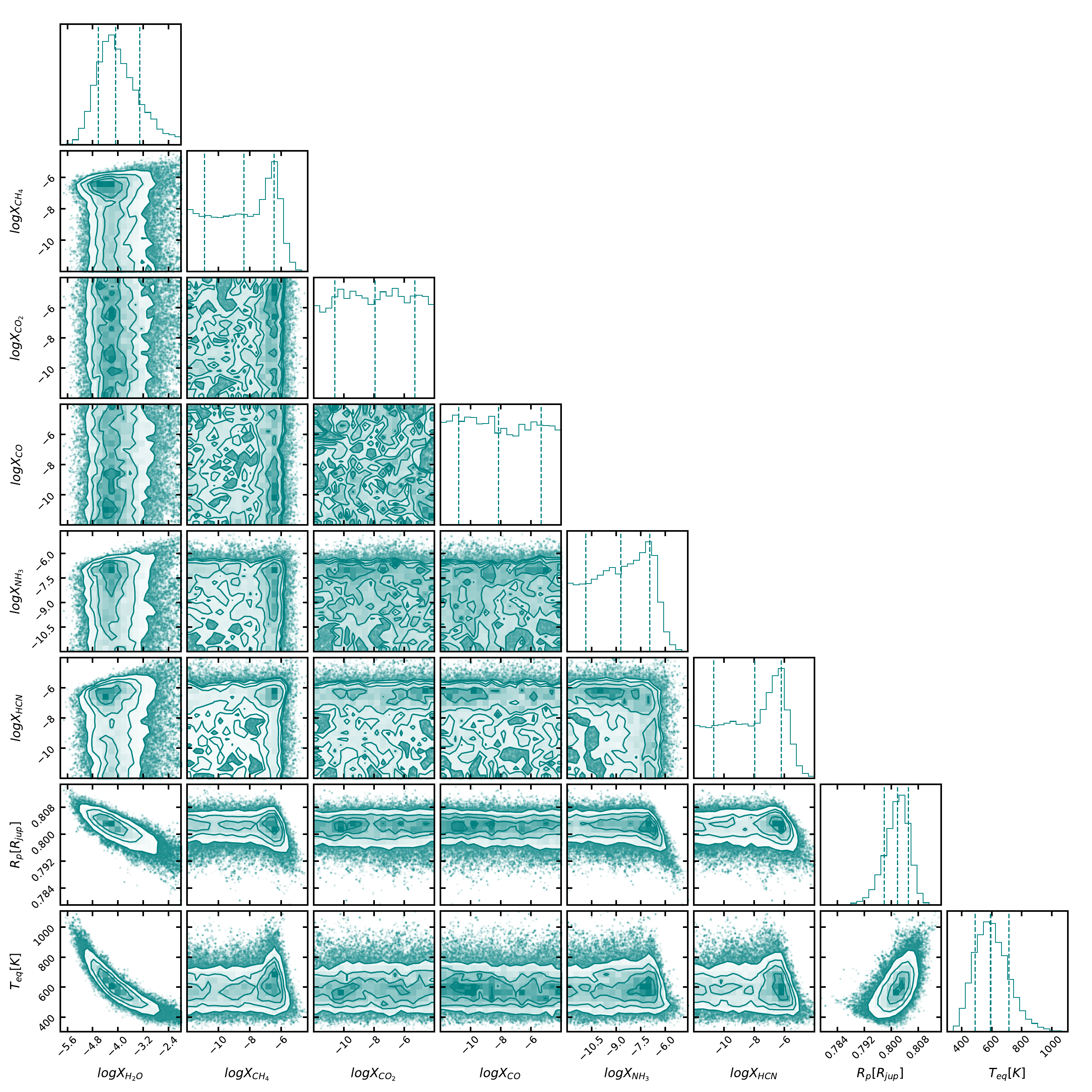}
    \caption{Posterior distribution from free atmopsheric retreival on uncorrected transmission spectrum of V1298 Tau b. We put strong constraints on water abundance and upper limits on CH4, HCN and NH3 abundance.  }
    \label{fig:free retreival}
\end{figure}

\begin{figure}
    \centering
    \includegraphics[width=\linewidth]{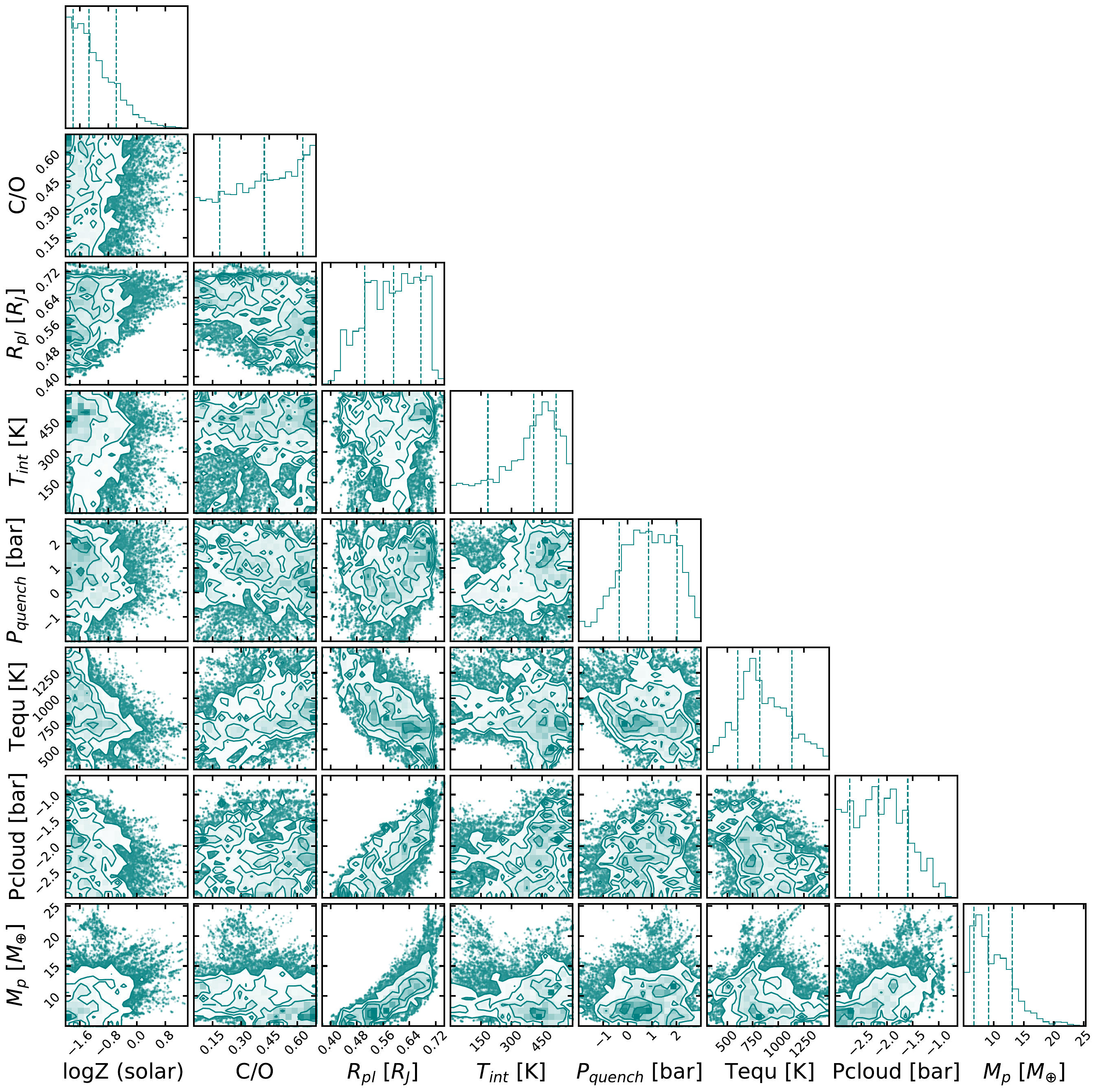}
    \caption{Posterior distribution of atmospheric retreival on the uncorrected transmission spectrum of V1298 Tau b where mass was not fixed. The mass posterior distribution peaks around 10$M_{\oplus}$ with an atmospheric metallicity 2$\sigma$ upper limit at solar metallicity. The posterior distribution of the planet mass is in agreement with the upper limit quoted in this paper.}
    \label{fig:free mass retreival}
\end{figure}

\begin{figure}
    \centering
    \includegraphics[width=\linewidth]{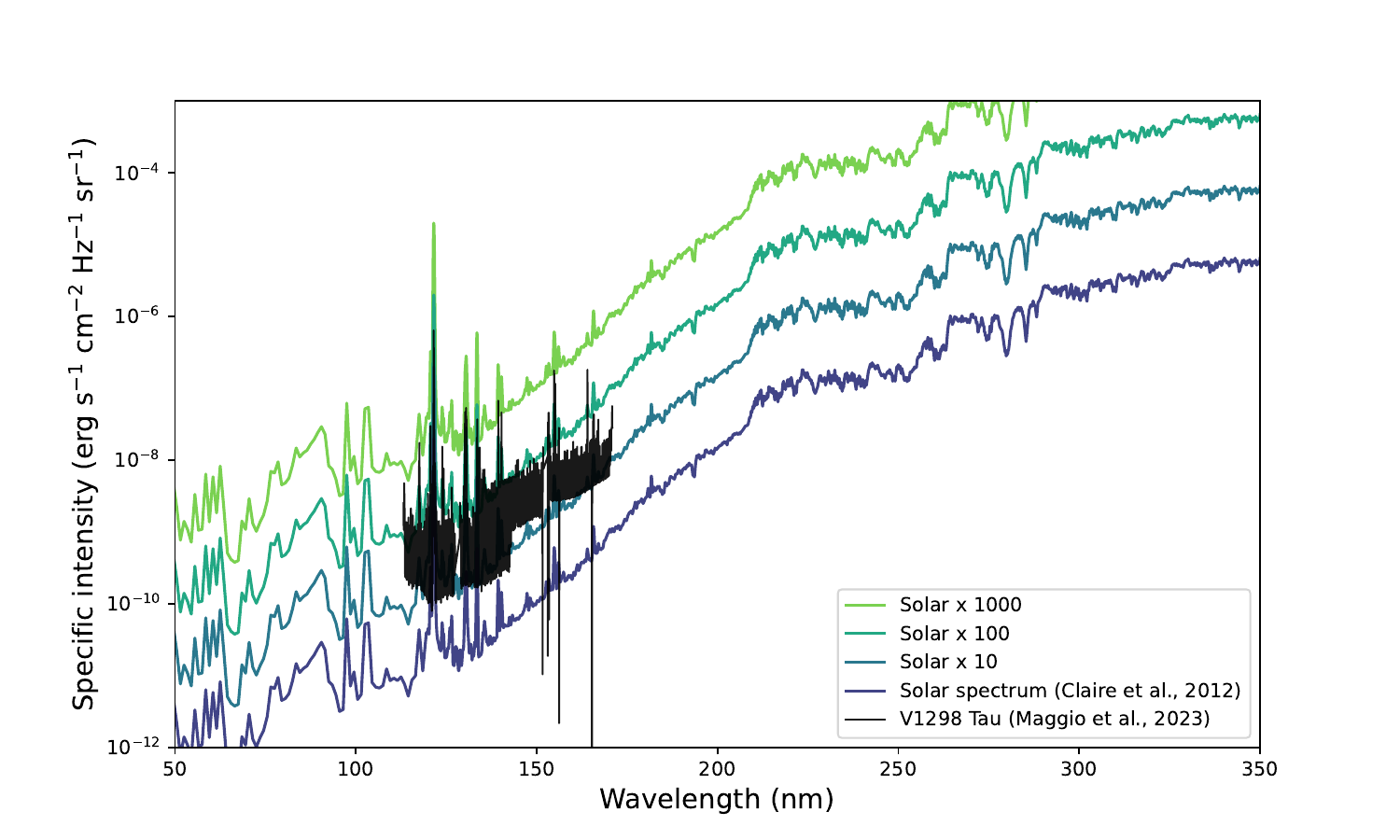}
    \caption{Spectral energy distribution of V1298 Tau compared to solar spectra.We see that the UV flux of the latter corresponds to that of the solar spectrum if it is scaled up between 10 - 100 times.  }
    \label{fig:stellar spectra}
\end{figure}

\begin{figure}
    \centering
    \includegraphics[width=\linewidth]{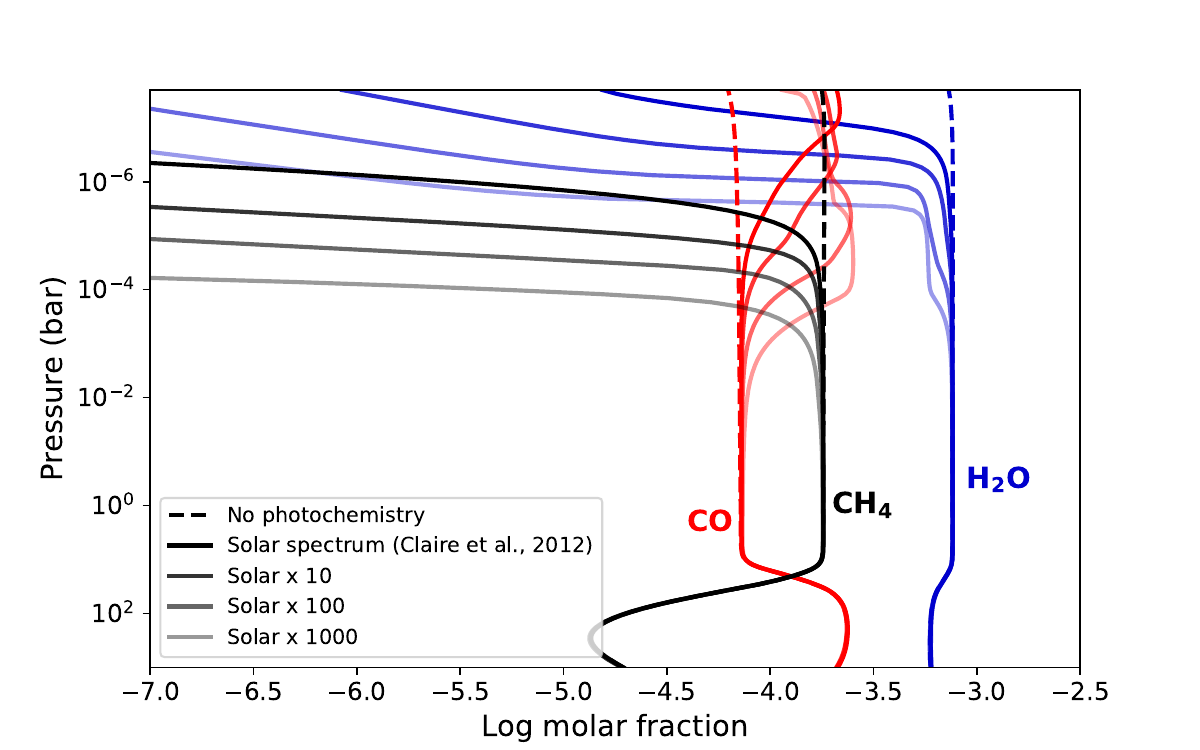}
    \caption{Self-consistent forward models using scaled solar solar spectra. With increasing UV flux, we find that water and methane start dissociating deeper in the atmosphere. However, for 10-100 times scaled solar spectra, which seems to represent V1298 Tau UV emission (Supplementary Figure \ref{fig:stellar spectra}), we find the effect of photochemistry to dominate for pressures lower than $10^{-4}$ bar. The models were setup identically to the ones presented in the paper. Thus, we can conclude that photochemistry is not likely to play a role in the removal of methane from the atmosphere of V1298 Tau b.}
    \label{fig:updated chemistry}
\end{figure}

\newpage
\begin{table}
\centering
\caption{Table showing the mass estimated using different radius reported in literature for V1298 Tau b}
\label{tab:mass_radius_comparison}
\begin{tabular}{c | c | c | c | c}
\hline
\hline
        Parameter & TESS only   & K2 only   & K2+TESS+RV  & HST white light curve \\  \hline
        Radius [$R_{J}$] & 0.85$\pm$0.03  & 0.91$\pm$0.05 & 0.88$\pm$0.03 & 0.84$\pm$0.003\\
        Mass [$M_{\oplus}$] &24$\pm$5& 19$\pm$4 & 21$\pm$5 &   18$\pm$4\\

        \hline
        \hline
\end{tabular}
\end{table}
\end{document}